\documentclass[10pt,sigplan,letterpaper,twocolumn]{article}
\usepackage[10pt,nocopyright]{sigmin}

\usepackage{listings}
\usepackage{color}
\usepackage{makecell} 
\usepackage[hyphens]{url}                                         
\usepackage[breaklinks,colorlinks]{hyperref}                      
\usepackage[usenames,dvipsnames]{xcolor}                          
\hypersetup{citecolor=blue,linkcolor=blue}                        
\usepackage{amsmath,amsopn,amssymb,amsthm}                        
\usepackage{thmtools}
\usepackage{thmbox}
\usepackage[toc,page]{appendix}
\usepackage{subcaption}                                           
\usepackage{endnotes,microtype,xspace,fancyvrb,multirow} 
\usepackage{epigraph} 
\usepackage{diagbox} 
\usepackage[normalem]{ulem}

\usepackage{booktabs}      
\usepackage{tabularx}
\usepackage{array,underscore,relsize}                             
\usepackage[T1]{fontenc}                                          
\usepackage{times}                                                
\usepackage{fancyhdr,lastpage}                                    
\usepackage{enumitem}                                             
\usepackage[labelfont=bf,font=normal,skip=1pt]{caption}           
\usepackage[export]{adjustbox}                                    
\usepackage{breakurl}                                             
\usepackage{pifont}                                               
\usepackage{tablefootnote}                                        
\usepackage{bbding}                                        
\usepackage[linesnumbered,vlined,boxed,commentsnumbered]{algorithm2e}
\DontPrintSemicolon

\usepackage{algpseudocode}
\usepackage{amssymb}
\usepackage[normalem]{ulem}
\usepackage[hang,flushmargin]{footmisc}
\usepackage{textcomp}
\usepackage{graphicx}
\usepackage{authblk}
\usepackage{amsmath}

\usepackage{colortbl}
\definecolor{lightgray}{RGB}{236,236,236}
\definecolor{redcolor}{RGB}{234,51,35}
\definecolor{greencolor}{RGB}{80,177,51}
\definecolor{blackcolor}{RGB}{0,0,0}

\usepackage[compact]{titlesec}
\titleformat*{\section}{\large\bfseries}
\titleformat*{\subsection}{\normalsize\bfseries}
\titleformat*{\subsubsection}{\normalsize\bfseries}
\titlespacing{\section}{0pt}{3ex}{1ex}
\titlespacing{\subsection}{0pt}{2ex}{1ex}

\hypersetup{
  colorlinks,
  linkcolor={red!50!black},
  citecolor={blue!50!black},
  urlcolor={blue!80!black}
}

\newcommand{\sys}{\textsc{SpecActor}}
\newcommand{\company}{\textsc{Seed}}

\newcommand{\verl}{{veRL}}

\newcommand{\fig}[1]{Figure{~\ref{#1}}}

\newcommand{\alg}[1]{Algorithm{~\ref{#1}}}
\newcommand{\eg}{e.g.,~}
\newcommand{\ie}{i.e.,~}

\newcommand{\one}{\texttt{\uppercase\expandafter{\romannumeral1}}}
\newcommand{\two}{\texttt{\uppercase\expandafter{\romannumeral2}}}

\newcommand{\nospacestitle}[1]{\noindent{\bf #1}\,}
\newcommand{\stitle}[1]{\vspace{1.1ex}\noindent{\bf #1}}
\usepackage{soul} 

\newcommand{\etitle}[1]{\vspace{0.5ex}\noindent{\em \ul{#1}}}

\usepackage{tikz}

\usepackage{balance}
\begin{document}
\RestyleAlgo{ruled}


\title{\Large \bf{
  Fast LLM Post-training via Decoupled and Fastest-of-N Speculation
}}

\author[1,2]{Rongxin Cheng}
\author[2]{Kai Zhou}
\author[1]{Xingda Wei\thanks{Xingda Wei is the corresponding author (\url{wxdwfc@sjtu.edu.cn}).}}
\author[1]{Siyuan Liu}
\author[1,2]{Mingcong Han}
\author[3]{Mingjing Ai}
\author[2]{Yeju Zhou}
\author[2]{Baoquan Zhong}
\author[2]{Wencong Xiao}
\author[1]{Rong Chen}
\author[1]{Haibo Chen}

\affil[1]{\vspace{1mm}Institute of Parallel and Distributed Systems, Shanghai Jiao Tong University\vspace{0.5mm}}
\affil[2]{ByteDance Seed\vspace{0.5mm}}
\affil[3]{Unaffiliated\vspace{-1.mm}}

\maketitle


\begin{abstract}
    \noindent
    Rollout dominates the training time in large language model (LLM) post-training,
    where the trained model is used to generate tokens given a batch of prompts.
    This work, {\sys}, achieves fast rollout with speculative decoding that deploys
    a fast draft path to accelerate the unparallelizable generation,
    while the correctness is guaranteed by fast parallel verification of the outputs with the original model.
    {\sys} addresses two foundational challenges that hinder 
    speculation efficiency:
    (1) a \emph{Decoupled speculation} method
    that overcomes the computation inefficiency issue when executing 
    speculative decoding with relative large per-worker batch size---
    a common configuration in training but unfriendly to speculation, 
    and (2) a \emph{Fastest-of-N speculation} method that selects and combines different draft methods
    according to the rollout progress to approximate the optimal draft method
    even when the best one is unknown a priori. 
    Extensive evaluations on production traces
    show that {\sys} accelerates mean rollout speed by {2.0--2.4}\,$\times$, 
    with up to {2.7}\,$\times$ speedup, over common post-training baselines.
    The results are consistent across both dense and MoE models and across different RL algorithms.
	Notably, {\sys} is {1.1--2.6}\,$\times$ faster
	compared to vanilla speculative rollout in different traces. 
    The accelerated rollout achieves {1.4--2.3}\,$\times$ faster end-to-end training time.

 \end{abstract}

\section{Introduction}
\label{sec:intro}

\noindent
Large foundational models like LLMs have demonstrated remarkable
accuracy across a wide range of tasks including
writing~\cite{DBLP:journals/corr/abs-2303-08774}, coding~\cite{DBLP:journals/corr/abs-2308-12950}, 
and many others~\cite{DBLP:journals/corr/abs-2112-09332,perf-agent,DBLP:journals/corr/abs-2504-05738,DBLP:journals/corr/abs-2502-01450}. 
Post-training these models with reinforcement learning (RL)
has become a key pillar in the training pipeline
to further enhance model capabilities 
in challenging reasoning tasks~\cite{DBLP:journals/corr/abs-2501-12948}.

Rollout is a key and performance-dominant phase in post-training (\textsection{\ref{sec:bg-motivation}}):
at each step,
the system feeds the model with a batch of prompts representing
target problems (e.g., math or coding problems) to be solved.
The model then generates tokens for each prompt in an attempt
to solve them.
When rollout finishes, the generated tokens are evaluated
and the model is updated accordingly.
Rollout naturally parallelizes the batch across multiple workers.

\stitle{Problem statement and current solutions (\textsection{\ref{sec:bg-motivation}}). \,}
Rollout accounts for {75--80}\% of the post-training time.
This lengthy and inefficient execution is primarily due to 
long tails in generation and is further exacerbated by idle GPU time, 
caused by waiting for the slowest worker to finish.
The long tail stems from difficulty variation, with some problems requiring more tokens than others to solve.
Therefore, one worker may finish its assigned prompts quickly,
while another takes much longer to complete.
Importantly, waiting for all requests before updating the model
is necessary to ensure rapid and stable training convergence 
~\cite{DBLP:journals/corr/SchulmanWDRK17,DBLP:journals/corr/abs-2402-03300,dapo}. 

Given this hard constraint, we ask a specific but important question:
\emph{how to accelerate rollout in LLM post-training {without changing
how the model is trained}?}

Existing \emph{algorithm-agnostic} solutions fall into two categories,
neither of which adequately addresses the issue 
given the recent trend of ever-growing generation length as
LLMs are trained to solve more complex problems~\cite{swebench,agentbench,deepseek-r1}.
The first category is \emph{overlapping}, which utilizes idle GPUs in the rollout for other
phases in the training pipeline~\cite{rlhfuse}.
It has been proven effective in short-context RL training but 
less effective in long-tailed training 
because it does not accelerate rollout directly, yielding 
1.1\,$\times$ average speedup.
The second category leverages \emph{parallelism},
\ie{utilizing idle or over-provisioned GPUs} to accelerate post-training~\cite{rlboost}.
Unfortunately,
the speedup is still limited (e.g., up to 1.2\,$\times$ with doubled GPUs)
because LLM generation is memory-bound and thus cannot fully utilize the extra computation power. 

\stitle{This work: lossless and efficient speculative rollout. \,}We present {\sys}, a fast rollout system that retrofits speculative decoding~\cite{DBLP:conf/osdi/BehrensCSBKZ20,DBLP:conf/osdi/ChangG99,spec-action,spec-edge}---a
common technique for accelerating LLM inference---to LLM post-training. 
Specifically, speculative decoding employs a fast path for sequential generation (decoding):
First, a ``draft'' model much smaller than the trained one 
generates a sequence of $n$ draft tokens.
Then, the trained model verifies these tokens. 
Once accepted, these tokens are treated as generated by the trained model.
The rationale is that
verifying $n$ tokens is {typically faster than generating them}---although 
using the same model---because
{verification processes multiple tokens in parallel.}

It is important to note that speculative decoding can provide lossless generation 
as the original decoding via exact token matching~\cite{DBLP:conf/acl/XiaYDW00L0S24}.
Moreover, the drafters are available at post-training~\cite{spd,specinfer}:
pre-training pipeline generates both small and large models,
following established scaling practices~\cite{qwen2.5,qwen3}.
Besides, n-gram-based drafters are model-free~\cite{pld,DBLP:conf/acl/HuWZZLCZ25}.

While intuitive and effective in inference, applying vanilla speculative decoding
to rollout yields only {10\,\%} speedup at worst on common production configurations 
(\textsection{\ref{sec:eval-e2e}})
due to the following challenges. We further address them with 
two contributions.

\stitle{Computation-efficient rollout via decoupled speculation (\textsection{\ref{sec:design-decouple}}). \,}
First, the effectiveness of speculative decoding depends mainly on the verification overhead,
which is proportional to the per-worker batch size. 
However, on common training configurations, e.g., per-worker batch size 128, (see {\fig{fig:motiv-batchsize}} (a)),
the verification could become slower than generation 
due to reaching the GPU computation limit,
also noticed by recent works~\cite{eagle-3,turbospec}.
Although the batch size gradually decreases during rollout as prompts finish,
it remains relatively large (\eg{$\geq$ 64}) for a considerable portion of the rollout time (\eg{20\,\%}),
so we need efficient execution on batch configurations 
that are unfriendly to speculation.

Our key observation is that there exist opportunities to accelerate verification because
vanilla speculative execution limits the GPU time provisioned to (compute-hungry) verification
due to a coupled draft-$n$-then-verify dependency:
the verifier must wait for GPU-underutilized drafter jobs.
Such a dependency is necessary to ensure high speculation efficiency for \emph{all requests} 
(required in inference~\cite{turbospec})
because without it, 
the drafter could further waste all $[n,...)$ tokens if verification rejects one token in $[1,n]$.
However, such a dependency is overly conservative for rollout 
where only the batch completion time matters,
so trading off a few requests to accelerate others is beneficial.
Based on this, we propose decoupled speculation to relax the dependency 
between draft and verification. 
The drafter can continue drafting without waiting for verification to finish,
thereby allowing subsequent verifications to start earlier to maximize GPU time provisioned to verification.
The decoupling comes at the cost of decreased acceptance rate due to more in-flight unverified tokens.
Fortunately, many requests only have a slightly decreased acceptance rate,
so the computation gain offsets the acceptance loss.
Moreover, we further propose a
feedback-based draft window mechanism to dynamically minimize wasted tokens due to such aggressive drafting 
via online reconfiguration.

\stitle{Effective draft adaptation via Fastest-of-N speculation (\textsection{\ref{sec:design-bon}}). \,}
Second, speculative decoding has multiple candidate draft methods,
and it is a common practice to select 
a single drafting method for all requests based on profiling or experience~\cite{spd,specinfer}.
However, we found in rollout that different requests suit different draft methods, 
so using one fixed method might lead to inefficient speculation 
for stragglers. 
However,
it is challenging to select the optimal one for all requests because
for each request, the optimal method not only depends on how fast the drafter is,
but also on the acceptance rate of verification.
The latter is unknown in advance as it depends on the model output. 
Meanwhile, it is impractical to run multiple draft methods for each request
as it requires proportional extra verifiers and thus extra GPUs.

We propose Fastest-of-N speculation that approximates the 
best draft methods for all requests with the following three techniques. 
First, we propose a new abstraction called \emph{draft ladder} that 
establishes a mapping from various acceptance rates to speedup given a pool of draft methods,
which can be constructed offline. 
Given the ladder, at the beginning of rollout,
we select one estimated fastest {draft method}
using the average acceptance rate of all requests profiled so far.
The rationale is that even though each requests' acceptance rate varies significantly,
the average acceptance rate over a (large) batch of requests is statistically stable~\cite{hoeffding1963probability}.
Thus, the selected draft method is likely to be close to the optimal for the batch. 
Finally, when the batch size decreases during rollout and
the initial selection becomes suboptimal,
we utilize the freed GPUs from finished workers to launch more draft methods in parallel for long-tailed requests,
and the request finishes when the fastest draft method 
generates the end token (EOS) that is accepted by the verifier.
Running multiple draft methods in parallel increases
the likelihood of selecting the fastest method
with unknown acceptance rates, while utilizing the idle GPUs
avoids extra GPU usage.

\stitle{Demonstration. }
{\sys} preserves the exact rollout process for any training algorithm,
so algorithm designers can seamlessly use it without worrying about potential accuracy loss.
We built {\sys} on {veRL}~\cite{hybridflow}, a state-of-the-art
post-training framework. 
{\sys} is easy to integrate into other training frameworks, as it serves as a drop-in replacement for the inference component 
of any post-training pipeline, without requiring changes to the training logic.
Extensive evaluations show that on various real-world training traces 
and common post-training algorithms including 
{GRPO~\cite{DBLP:journals/corr/abs-2402-03300,deepseek-r1}, DAPO~\cite{dapo}, and PPO~\cite{DBLP:journals/corr/SchulmanWDRK17}} , 
compared to the state-of-the-art solutions like {veRL and RLHFuse},
{\sys} accelerates the end-to-end training time by {1.4--2.3}\,$\times$.
Also, we are up to {2.6}\,$\times$
{faster than vanilla speculative rollout baselines in different traces.}

\section{Background and Motivation}
\label{sec:bg}

\subsection{LLM post-training}
\label{sec:bg-post-training}

\begin{figure}[!t]
        \begin{minipage}{1\linewidth}
        \centering    
        \includegraphics[width=1\linewidth, trim=0.25cm 10.95cm 19.4cm 0.25cm, clip]{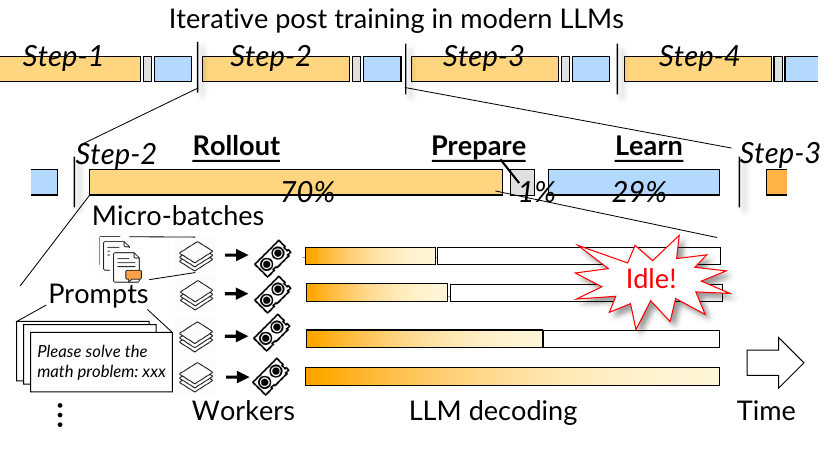} 
        \end{minipage} \\[0pt]
        \begin{minipage}{1\linewidth}
        \caption{\small{%
            An illustration of the rollout process in LLM post-training. 
        }}
        \label{fig:bg-rollout}
        \end{minipage} \\[-20pt]
        \end{figure} 

\nospacestitle{Generation with LLM. \,}
Large Language Models (LLMs) generate responses (termed \emph{tokens}) through \emph{prefill} and \emph{decode} phases.
The prefill phase processes the initial input (termed \emph{prompt})
 to generate the first token.
The decode phase then iteratively generates subsequent tokens in an auto-regressive fashion
by combining previous output tokens with the original input as the new input.
The decode phase terminates upon generating an end-of-sequence token.

\stitle{Post-training workflow. \,}
Leading LLMs are post-trained with reinforcement learning (RL) 
with multiple steps of execution,
where each step follows an identical three phases~\cite{DBLP:journals/corr/abs-2501-12948,realhf,DBLP:conf/nsdi/ZhongZWLCWHXMZ025,hybridflow}: 

\etitle{1. Rollout. \,}
The process treats the trained model as actor and generates tokens
from a sampled batch of pre-defined prompts using LLM decoding.
The tokens are used to solve specific tasks like math problems. 
To accelerate rollout, the sampled batch is divided into smaller micro-batches 
(\emph{batch}\footnote{\footnotesize{Since we only refer to micro-batches in the following content,we use the more general term \emph{batch} to indicate micro-batch without loss of generality. }})
and processed concurrently 
by multiple rollout workers. 
Each worker deploys the model on one or more GPUs 
using model parallelism if necessary~\cite{hybridflow}.

Two aspects regarding rollout batch need to be noted.
First, to improve efficiency, production RL training applies a relatively large
rollout batch size per step to find sufficient positive rewards~\cite{DBLP:conf/iclr/KeskarMNST17,DBLP:journals/corr/abs-2402-03300,DBLP:conf/nips/YaoGLKM18,dapo,gspo},
e.g., a typical 32B task
sets the per-step total batch size to {8\,K}~\cite{dapo,DBLP:journals/corr/abs-2504-05118}. 
Second, each prompt is typically seen only once or twice 
(\ie{common training involves just 1--2 epochs~\cite{deepseek-r1}}),
limiting the effectiveness of n-gram-based methods like experience replay
~\cite{liu2025specrlacceleratingonpolicyreinforcement,rhymerl}.

\etitle{2. Prepare. \,} 
In the prepare phase, the rollout results are fed into
a set of judgers~\cite{DBLP:journals/corr/abs-2402-03300,dapo,DBLP:journals/corr/abs-2504-02605,DBLP:journals/corr/abs-2509-02547}
(e.g., a separate reward model)
to generate rewards~\cite{DBLP:journals/corr/SchulmanWDRK17,DBLP:journals/corr/abs-2204-05862},
which are the arithmetic signals that guide the parameter optimization.
These judgers are lightweight; for example, the reward models only compute a forward pass,
rather than auto-regressive generation,
so the time required is negligible in a training step.

\etitle{3. Learn. \,}
Given the signals from the prepare phase, the learning phase 
calculates the loss and updates the LLM parameters
with a backward pass.
The updated parameters are then used in the next rollout step. 

\begin{figure}[!t]
        \begin{minipage}{1\linewidth}
        \hspace{-1mm}
        \centering    
        \includegraphics[width=\columnwidth]{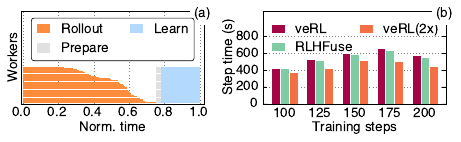} \\[1pt]    
        \end{minipage} \\[0pt]
        \begin{minipage}{1\linewidth}
        \caption{\small{%
        {
            (a) {The long rollout in post-training}
            and (b) the training latency of various steps
            in DAPO-32B-20K training trace (detailed setups described in \textsection{\ref{sec:eval-setup}}). 
        }}}
        \label{fig:data-motivation}
        \end{minipage} \\[-5pt]
\end{figure}

\subsection{Analysis of post-training and current solutions}
\label{sec:bg-motivation}

\begin{figure}[!t]
        \begin{minipage}{1\linewidth}
        \centering    
        \includegraphics[width=1\linewidth, trim=0.25cm 12.78cm 18.75cm 0.25cm, clip]{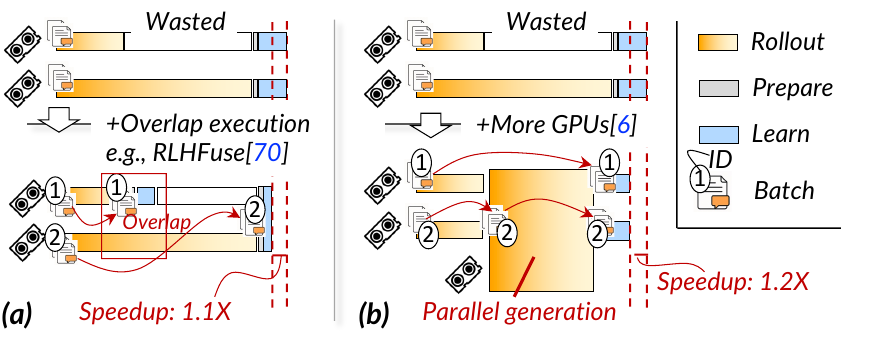} 
        \end{minipage} \\[0pt]
        \begin{minipage}{1\linewidth}
        \caption{\small{%
            (a) An illustration of accelerating post-training via overlapped execution and 
            (b) an illustration of accelerating rollout through scaling to more GPUs.
        }}
        \label{fig:motiv-current-solution}
        \end{minipage} \\[-15pt]
        \end{figure}

\nospacestitle{Rollout dominates execution time in post-training and suffers significant GPU wastes. \,}
{\fig{fig:data-motivation}} (a) analyzes the contribution of total post-training time in 
typical training tasks:
rollout contributes 70--80\,\% of the total training time.
Since post-training could last for days on hundreds of thousands of GPUs,
accelerating rollout is critical to reducing the overall training cost. 

There exists significant space for improvement for rollout 
because many GPUs are idle during the rollout phase due to the long-generation tail problem.
{\fig{fig:data-motivation}} (a) shows the GPU bubble in long-tailed rollout:
on a {DAPO-32B-20K} trace, we observed averagely {50\,\%} of the total GPU time is wasted
on waiting for the slowest worker to finish.
Rollout time differs significantly across workers
even with the same per-worker batch size,
because the number of tokens required to solve different tasks varies significantly,
and the exact number is hard to predict in advance as they are determined by the LLM outputs.
Moreover, the waste increases as training progresses because as the model becomes "smarter",
it tends to generate more tokens to solve particular tasks~\cite{deepseek-r1}.

\stitle{Reducing GPU waste via overlapping achieves a minor speedup for long-tail generation. \,}
An intuitive solution to reduce wasted GPU time is to overlap other phases with the current rollout:
As shown in {\fig{fig:motiv-current-solution} (a)},
representative solutions like RLHFuse~\cite{rlhfuse} utilize the idle workers
to execute the prepare and learn phases of finished batches (e.g., batch \ding{192}),
overlapping these phases from fast workers with the rollout of slower workers.
While such an overlap cannot accelerate the rollout phase, which is bottlenecked by the slowest worker,
it reduces the end-to-end training time
because once the slowest rollout finishes, more GPUs can participate in accelerating
the remaining phases of the slowest batch (e.g., batch \ding{193}).

However, the speedup from overlapping is limited
because the accelerated prepare and learn phases only contribute a small portion of the overall training time (\eg{about 5\% in {\fig{fig:data-motivation}} (a)}).
As a result, RLHFuse only accelerates post-training by {3\,\%} in traces with long response budget,
with GPU time still wasted, as shown in {\fig{fig:data-motivation}} (b).

\stitle{The sequential-generation nature of rollout is the key obstacle for acceleration. \,}
Another intuitive solution is to dynamically add more computation resources to accelerate the long-tailed workers:
as shown in {\fig{fig:motiv-current-solution} (b)},
once a rollout worker finishes its assigned batch (e.g., batch \ding{192}),
it can help accelerate the generation of unfinished batches (e.g., batch \ding{193}) using
parallel LLM generation techniques like sequence parallelism~\cite{loongserve}.
Moreover, current works~\cite{rlboost} aggressively add more GPUs (e.g., by allocating spot instances)
to accelerate the slow rollout workers without waiting for more GPUs to become idle.
In our example, three workers now process the rollout of batch \ding{193}.
Unfortunately, 
as shown in {\fig{fig:data-motivation}} (b), even if we provision 2\,$\times$ GPUs for rollout, 
the end-to-end training time is only improved by {1.2--1.3\,$\times$}.
The limited speedup is rooted in the 
sequential, memory-bound nature of rollout: 
tokens must be generated one-by-one, 
leaving little room for tensor or sequence parallelism~\cite{loongserve} 
due to the extra communication overhead.
Moreover, data and pipeline parallelism can hardly accelerate 
when memory-bound.

\vspace{2.mm}
\noindent \fbox{
\parbox{0.96\linewidth}{
\noindent {Current systems either suffer from limited speedup or adopt lossy acceleration 
methods like off-policy training or truncating tailed generations,
as summarized in \textsection{\ref{sec:related-work}}. 
We seek an algorithm-agnostic system solution for fast rollout. 
}}
}

\section{Design Rationale and System Overview}
\label{sec:overview}

\begin{figure}[!t]
       \vspace{2mm}
        \begin{minipage}{1\linewidth}
        \centering    
        \includegraphics[width=0.95\linewidth, trim=0.25cm 22.4cm 26.6cm 0.25cm, clip]{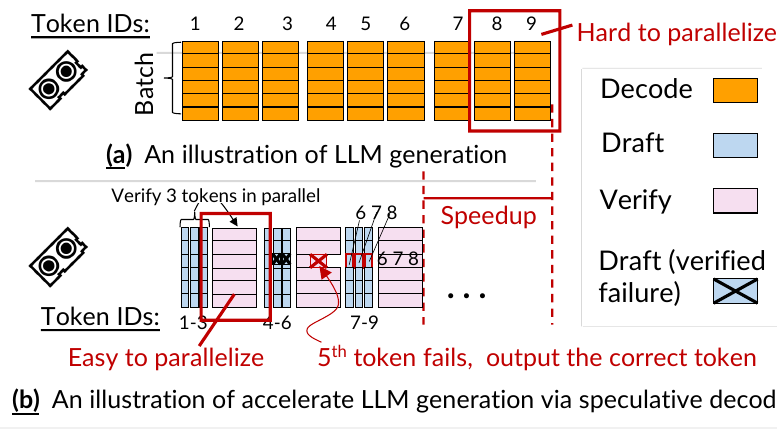} 
        \end{minipage} \\[2pt]
        \begin{minipage}{1\linewidth}
        \caption{\small{%
        An illustration of speculative decoding.  
        }}
        \label{fig:spec-decoding}
        \end{minipage} \\[-15pt]
        \end{figure} 

\nospacestitle{Opportunity: speculative decoding for rollout}~\cite{DBLP:journals/corr/abs-2302-01318,DBLP:conf/icml/LeviathanKM23,eagle-3,turbospec,DBLP:conf/nips/SternSU18,DBLP:conf/asplos/MiaoOZCWZWZYSSC24}.
It is a common technique to accelerate sequential LLM generation:
As detailed in \fig{fig:spec-decoding} (a),
the original generation iteratively generates tokens for the current batch
using the model being trained. 
With speculative decoding (b),
we first generate a sequence of tokens using a fast draft method,
e.g., with a smaller model, as detailed in \textsection{\ref{sec:design-decouple}}.
The drafted tokens are then verified by the trained model 
before being accepted as the generated tokens.

\begin{figure}[!t]
        \vspace{4mm}
        \begin{minipage}{1\linewidth}
        \centering    
        \includegraphics[width=0.9\linewidth, trim=0.25cm 18.4cm 20.35cm 0.25cm, clip]{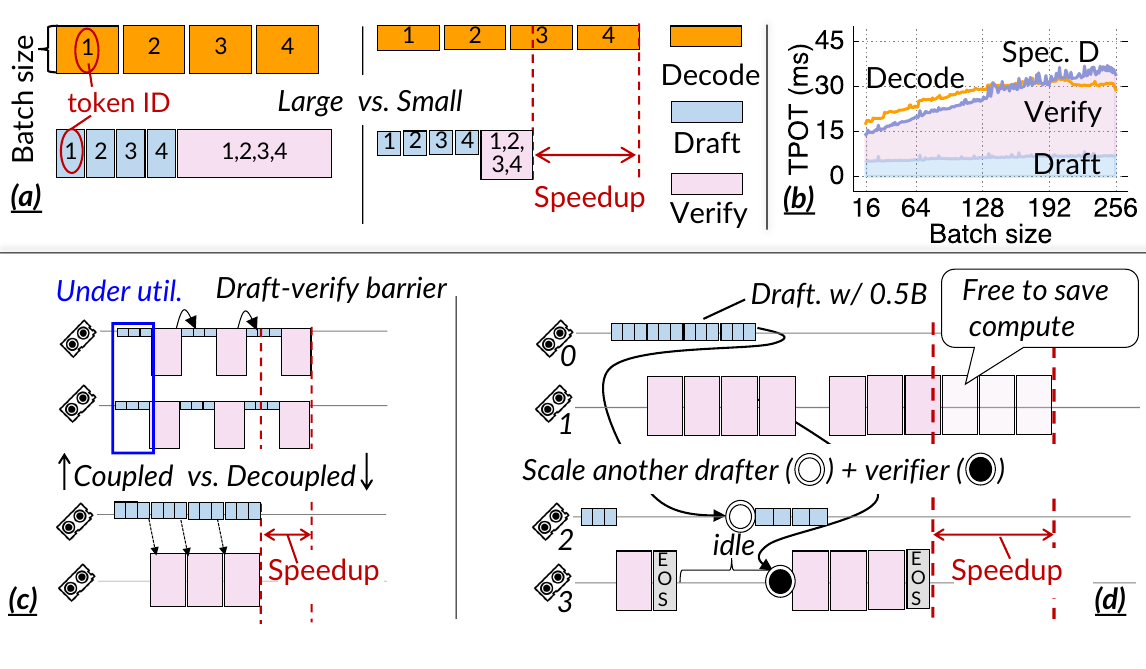}
        \end{minipage} \\[1pt]
        \begin{minipage}{1\linewidth}
        \caption{\small{%
        (a) An illustration of the limited speedup of speculative decoding with
        increased verification costs due to increased batch size.
        (b) A TPOT (Time Per Output Token) analysis of speculative and
        normal decoding of Qwen2.5-32B with different batch sizes.
        (c) An illustration of how decoupled execution reduces GPU underutilization
        due to the drafter.
        (d) An illustration of how Fastest-of-N speculation works with finished batches.
        }}
        \label{fig:rollout-schedule}
        \end{minipage}
        \end{figure}

Since the verification can be performed on multiple tokens in parallel,
its time is similar to or less than that of generating a single token if GPU computation is not the bottleneck. 
Thus, if tokens are accepted, speculative decoding is much faster than the original generation,
as shown in the right half of \fig{fig:rollout-schedule} (b).
On the other hand, if the verification rejects the drafted tokens,
e.g., the 5$^{th}$ token for request 3 in our example,
the rejected token and draft tokens after it,
e.g., tokens 5--6, will be discarded, and the drafter starts
from token 6---as the verification will provide a correct 5$^{th}$ token.

\begin{figure}[!t]
        \vspace{2mm}
        \begin{minipage}{1\linewidth}
        \centering    
        \includegraphics[width=1\columnwidth,trim=0.25cm 24.37cm 24.8cm 0.25cm, clip]{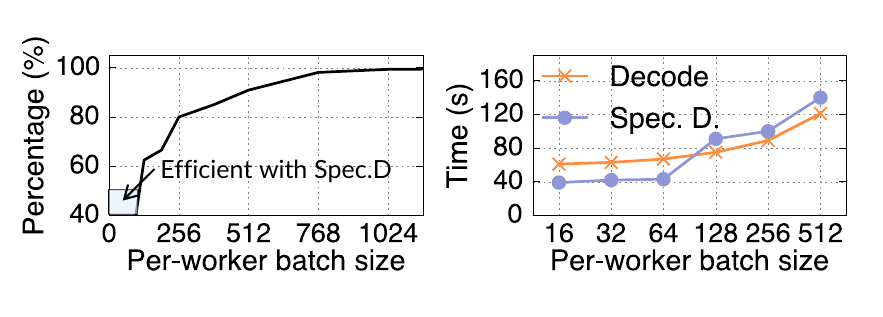} \\[3pt]
        \end{minipage} \\[0pt]
        \begin{minipage}{1\linewidth}
        \caption{\small{%
        {
            (a) Distribution of the initial per-worker batch sizes of post-training traces in the last 6 
            months in a large production cluster. 
            (b) The acceleration of speculative rollout given such a batch size
            on a {Qwen2.5-32B} checkpoint. 
        }}}
        \label{fig:motiv-batchsize}
        \end{minipage} \\[-5pt]
\end{figure}

\stitle{Challenge \#1: Limited speedup for typical training configurations. \,}
While intuitive, we found adopting state-of-the-art speculative decoding techniques\cite{pld,lookahead,eagle-2,specinfer}
in rollout results in limited speedup,
because the speculation verification is inefficient on 
per-worker batch sizes common in training (see {\fig{fig:rollout-schedule}} (a)). 
{\fig{fig:motiv-batchsize}} (a) shows the distribution of per-worker batch sizes collected
from production post-training jobs, and 
(b) compares the time of generation at most {4,096} tokens 
using speculative decoding and the original generation
on a {Qwen2.5-32B} model.
We can see that for common per-worker batch sizes (e.g., 128), speculative decoding brings no
or negative gain.
The reason is that the verification time increases more significantly 
with the increased request batch size than others,
as shown in {\fig{fig:rollout-schedule}} (d), 
because verification introduces a larger token batch 
than original generation.

One may notice that the per-worker batch size decreases during rollout 
as more requests finish.
However, in at least 20\% of the rollout time, 
the batch size is relatively large
in our DAPO-32B-20K trace (see \textsection{\ref{sec:eval-setup}}),
which still calls for efficient speculative decoding acceleration.
In training, even a tiny acceleration matters due to the
potential of saving many GPU hours.

\etitle{Our solution: decoupled speculation (\textsection{\ref{sec:design-decouple}}). \,}
As we have mentioned in the introduction, we decouple the dependency between
drafting and verification,
which allows us to fully utilize GPUs for speculative decoding.
As shown in {\fig{fig:rollout-schedule}} (c),
unlike traditional coupled execution where the drafter must wait for the verifier,
we (1) distribute the drafter and verifier to separate GPUs and
(2) allow the drafter to aggressively continue without waiting for the verifier.
Such an execution scheme provides more GPU time to the verifier
because we only need to allocate a few GPUs for the drafter
to avoid many GPU executions being GPU-underutilized during drafting.
Note that for the same set of GPUs, decoupled execution increases
the per-worker batch size for the verifier:
in our example, the batch size is doubled.
This does not offset the gain brought by decoupling because
verification with a 2\,$\times$ batch (from 128 to 256)
only incurs a 1.4\,$\times$ higher latency (as shown in {\fig{fig:rollout-schedule}} (b)).
Besides, our placement method further minimizes the cost by configuring
an appropriate parallelism that 
distributes the verification across more GPUs.

Decoupled speculative decoding faces the challenge of wasting more GPU resources upon 
possible decreased acceptance rates. 
In the example in {\fig{fig:rollout-schedule}} (c),
if a token in positions 1--3 is mis-speculated, extra drafted tokens (4--6) will be discarded,
which could not happen in a coupled speculation. 
Fortunately, we found it has little impact on many requests since their
acceptance rates remain high, so the speedup of decoupled execution offsets the loss due to waste,
still enabling a rapid reduction in the per-worker batch size.
To further cope with requests that suffer from significantly decreased acceptance rates,
we only relax but not fully discard the draft-verify dependency, 
using a draft window to control the aggressive generation,
i.e., the drafter is only allowed to be ahead of the verifier by a certain number of tokens.
Our per-request reconfiguration effectively adjusts these windows online
as long as we detect a significant decrease in the acceptance rate for a request.

\begin{figure}[!t]
        \begin{minipage}{1\linewidth}
        \hspace{-1mm}
        \centering    
        \includegraphics[width=0.85\columnwidth,trim=0.25cm 21.2cm 29.8cm 0.25cm, clip]{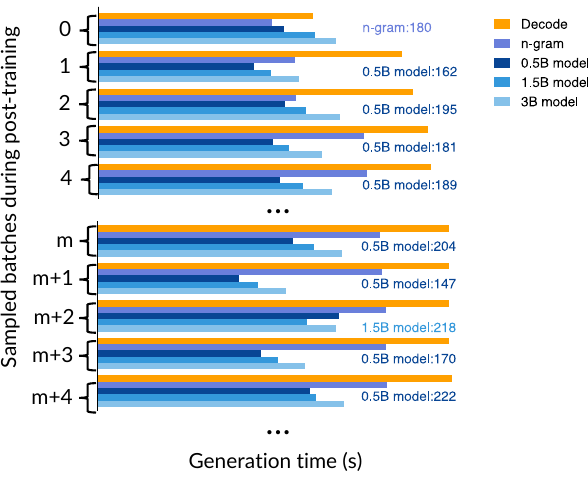} \\[3pt]
        \end{minipage} \\[-5pt]
        \begin{minipage}{1\linewidth}
        \caption{\small{%
        {
                A characterzation of the speedup of different draft methods 
                on DAPO-32B-20K (see \textsection{\ref{sec:eval-setup}}) training trace. 
        }}}
        \label{fig:motiv-bon}
        \end{minipage} \\[-15pt]
\end{figure}

\stitle{Challenge \#2: Selecting the best draft methods without priori knowledge (\textsection{\ref{sec:design-bon}}).  \,}
Selecting the best draft method is critical to ensuring high speedup yet is quite challenging
because the optimal draft method for each prompt varies and is unknown a priori.
The diversity stems from the variation in acceptance rates for a given request.
{\fig{fig:motiv-bon}} illustrates the speedup of different draft methods
on different requests from a batch collected from the training trace.
We can see that although many requests achieve the highest speedup with 0.5B drafting models,
some require 1.5B models and others require statistical methods like n-gram.
Directly executing multiple draft methods in parallel is infeasible
because it requires significantly more GPUs---not only to host multiple draft models but also to verify their outputs.

\etitle{Our solution: Fastest-of-N speculation. (\textsection{\ref{sec:design-bon}})\,}
At the start of the rollout, 
we use the profiled acceptance rate to select the estimated best draft method,
based on the observation that although each prompt's acceptance rate varies significantly,
for a large batch, 
the average acceptance rate is statistically stable.
Thus, the selection based on the average is likely to be close to the optimal for the batch.
The best draft is selected based on a ladder we constructed offline, 
which holistically considers major methods, 
including model-based~\cite{spd} and n-gram-based~\cite{pld,lookahead}.

During the rollout, when more GPUs are freed due to finished requests,
{\sys} dynamically utilizes freed GPUs to deploy more draft methods for tailed requests,
where parallel drafting naturally approximates the best draft method better,
because as long as the fastest draft method generates all the tokens, 
we can treat the requests as finished. 
As shown in {\fig{fig:rollout-schedule}} (d),
if worker 3 becomes idle, we add the second-best draft method (1.5B)
to accelerate the unfinished requests.
Note that in the illustration we only add one more verification instance
since the drafter can be piggybacked on other workers as it is relatively lightweight.
If the 1.5B drafting model finishes earlier,
the request is completed, and the request is also removed from other workers (e.g., worker 0 and 1).

\begin{figure}[!t]
        \hspace{-1mm}
        \begin{minipage}{1\linewidth}
        \hspace{1mm}
        \includegraphics[width=0.95\linewidth, trim=0.25cm 12.02cm 25.63cm 0.25cm, clip]{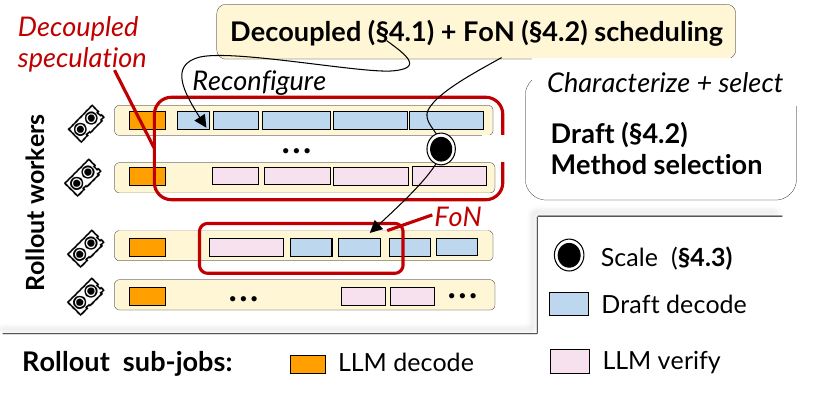} 
        \end{minipage} \\[2pt]
        \begin{minipage}{1\linewidth}
        \caption{\small{%
            The system architecture of {\sys}.
        }}
        \label{fig:syc-arch}
        \end{minipage} \\[-15pt]
        \end{figure}  

\stitle{System architecture. \,}
{\fig{fig:syc-arch}} shows the system architecture of {\sys}:
a global scheduler searches an efficient decoupled
speculation plan (\textsection{\ref{sec:design-decouple}}) during the initial
rollout phase.
Each worker dynamically reconfigures the 
execution plan for low-acceptance-rate requests at runtime to improve efficiency.
The global scheduler monitors the GPU usage across workers to deploy 
additional draft methods for long-tailed requests
(\textsection{\ref{sec:design-bon}}).
Our runtime (\textsection{\ref{sec:design-mechanisms}}) efficiently supports
scaling primitives required for Fastest-of-N speculation. 

\section{Detailed Design and Implementation}
\label{sec:design}

\subsection{Efficient Decoupled Speculative Rollout}
\label{sec:design-decouple}

\nospacestitle{Methodology. \,}
We adopt a two-step scheduling method to realize an efficient 
decoupled speculation:
(1) at the beginning of the rollout step,
we determine the optimal placement---how many workers are assigned to
drafting and verification, respectively---to minimize 
batch completion time.
During rollout, we (2) dynamically monitor the acceptance
rate of running requests to timely adjust their number of 
aggressively drafted tokens (with a draft window)
to minimize the impact of verification failures.

We choose to adjust only the draft window per request
but not the entire placement for the running batch during rollout for
three reasons. First, the optimal placement relies on
the modeling of system execution, but it becomes less
accurate due to the continuously reduced batch sizes caused by finished requests.
Second, the overhead of reconfiguring placement may outweigh the benefits.
Finally, since the batch sizes shrink quickly thanks to our decoupled execution
with optimal initial placement,
we no longer need to maximize computational efficiency,
and the fine-grained adjustment of the tail requests provided by (2)
enables sufficient speedups.

Below, we first describe how we control the wasted
tokens via \emph{draft window}, and then proceed to the two policies.

\begin{figure}[!t]
       \vspace{2mm}
        \begin{minipage}{1\linewidth}
        \centering    
        \includegraphics[width=0.95\linewidth, trim=0.25cm 24.0cm 26cm 0.25cm, clip]{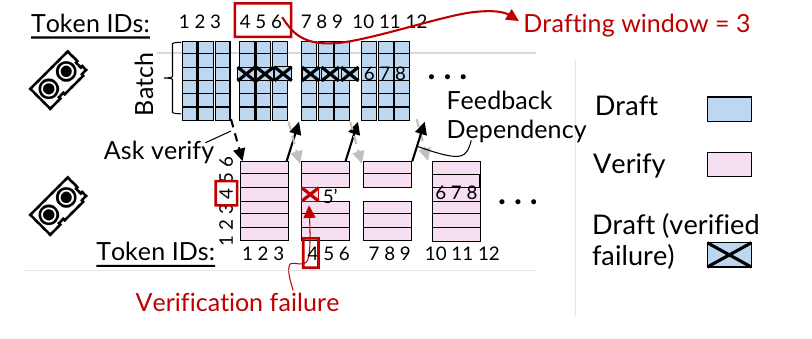} 
        \end{minipage} \\[0pt]
        \begin{minipage}{1\linewidth}
        \caption{\small{%
            An illustration of how relaxed decoupled execution 
            handle verification failure
            with a draft window of 3 tokens. 
        }}
        \label{fig:decoupled-cases}
        \end{minipage} \\[-15pt]
        \end{figure}

\stitle{Drafting window ($w$) and relaxed draft--verification dependency. \,}
To control the waste due to mis-speculation,
we set a window similar to previous works~\cite{turbospec} such that:
once the drafter sends $w$ tokens to the
verifier,
it is only allowed to aggressively draft another $w$ tokens.
If the verifier has not finished verification,
the drafter stops waiting for the feedback from the verifier. 
In this way, the drafter wastes at most $2w - 1$ tokens under speculation failures,
as shown in {\fig{fig:decoupled-cases}}.

\begin{figure}[!t]
       \vspace{2mm}
        \begin{minipage}{1\linewidth}
        \hspace{-1mm}
        \centering    
        \includegraphics[width=\columnwidth]{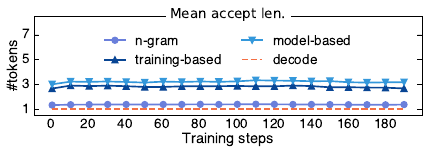} \\[3pt]    
        \end{minipage} \\[0pt]
        \begin{minipage}{1\linewidth}
        \caption{\small{%
        {
            Mean acceptance length profiled in a 200-step DAPO-32B-20K trace, covering 327\,K
            sequences. The training-based method uses \emph{frozen} EAGLE released 
            by TLT~\cite{tlt}.
        }}}
        \label{fig:motiv-acc-len}
        \end{minipage} \\[-10pt]
\end{figure}

\stitle{1. Decoupled speculation execution plan. \,}        
At the beginning of each rollout step,
given a draft method (described in \textsection{\ref{sec:design-bon}}),
the trained model, and the available GPUs,
our planner determines
an appropriate draft window for the entire batch, and
an efficient initial decoupled speculation plan by assigning
the drafter and verifier to different GPUs.
The plan only needs to be executed once during post-training
because for each step,
the initial batch size per worker remains the same,
and the acceptance length---average output length per verification 
are stable for mainstream drafters,
as shown in \fig{fig:motiv-acc-len}. 
Our reported EAGLE acceptance length\footnote{\footnotesize{No prompt tuning was applied following TLT authors' instructions.}} is lower than that 
of TLT~\cite{tlt} because our rollout
temperature is set to 1.0 and we use a 
large-batch-tuned configuration---a common production setup in post-training~\cite{deepseek-r1,DBLP:journals/corr/abs-2402-03300,DBLP:journals/corr/abs-2507-20534}.

Finding the optimal execution plan requires solving a combination
problem of the parallel configurations of different drafters and verifiers,
as well as precisely modeling the performance of speculative decoding using the draft window.
To ease problem formulation, we follow prior works that
assume the developers have provided a set of possible GPU configurations for running a verifier ($\mathbb{G}$),
i.e., how one copy of model parameter is partitioned on a set of GPUs. 
For drafter, we assume it only uses one GPU as it is lightweight. 

Algorithm~\ref{alg:argmax-select} describes our algorithm, 
which is essentially an enumeration-based search with
decoupled-execution-aware pruning to accelerate the process.
{First, given a batch of requests to generate (\ie{$B$}),
we select a verification configuration (line 2),
and estimate its performance under decoupled execution ($\mathrm{TGS}_{\text{cur}}$)
for various numbers of draft GPUs (line 3) and
various draft window sizes (line {6}).
We continuously select execution plans and finally choose the one with the
minimal estimated generation time (line 8--10).}

To improve search time, we prune the enumeration space by
(1) using the observation that drafters need fewer GPUs than verifiers (line {3})
and (2) pruning arbitrarily large draft windows,
because it only increases the computation waste upon mis-speculation (line {5}).

\stitle{Modeling $\mathrm{TGS}$. \,}
A key aspect of the planning is to model the performance---\uline{t}oken 
\uline{g}eneration \uline{s}peed ($\mathrm{TGS}$)---for 
decoupled execution. 
We use a bottom-up approach that first models the execution time of drafter 
and verifier, then move on the modeling of 
the expectation of $\mathrm{TGS}$,
considering the impact of wasted tokens due to mis-speculation
under a selected draft window.

Given a batch of request $b$, the number of drafter workers $g_d$,
and a verifier with execution configuration $g_v$,
the draft ($D_{g_d}(\cdot)$) and verification ($V_{g_v,w}(\cdot)$) time can be approximated 
as an affine function with respective to the batch size ($b$):  \\[-10pt]
\begin{align*}
    {D_{g_d}(b)} &= b \cdot D_{g_d}' + \alpha_{g_d} \\
    {V_{g_v,w}(b)} &= b \cdot V_{g_v,w}' + \beta_{g_v,w}
\end{align*}
\noindent
where $D_{g_d}'$, $V_{g_v,w}'$, $\alpha_g$ and $\beta_{g_v,w}$ are 
hyperparameters that fitted through offline profiling,
similar to prior works~\cite{DBLP:conf/osdi/ZhuGZZZGX0KLWWK25,kunserve}. 

Therefore, the total generation time of $w$ tokens in our decoupled execution paradigm
can be modeled as: \\[-10pt]
\begin{align*}
    \mathrm{IL}_{g_d,g_v,w}(b) = \max \{wD_{g_d}(b), V_{g_v,w}(b)\}    
\end{align*}

Besides the execution time, we also need to consider the 
slowdown due to mis-speculation. 
Given a draft window $w$ and the probability of accepting one token $p$ for a batch of
requests, we follow prior works~\cite{DBLP:journals/corr/abs-2505-19645,turbospec} that
first model the probability of accepting $\alpha$ tokens
given $n$ tokens within a draft window,
then calculate the expected wasted tokens accordingly.
The difference is that we further consider the additional wasted tokens due to
aggressive speculation in decoupled execution. 
Specifically, within a draft window, 
the probability of accepting $a$ tokens 
given an estimated accept probability ($p$) is: \\[-10pt]:
\begin{align*}
    \mathrm{P}(a, w) &=
\begin{cases}
p^a (1 - p), & 0 \leq a \leq w - 1,\\[0.5em]
p^a, & a = w,
\end{cases} \\
    s.t.  &\quad a \in \mathbb{N}, n \in \mathbb{Z^+}
\end{align*}
\noindent
$p$ is profiled using the setup from previous steps 
and we found it is quite stable
for a relatively large batch of requests 
(see \fig{fig:motiv-acc-len}).

The expectation of the total generated tokens for drafting a draft window $w$ is: \\[-10pt]
\begin{equation*}
    \tau_w =
\overbrace{\sum_{a=0}^{w-1} p^{a}(1-p)
        \frac{a+1}{2}}^{\text{partial accept}}
+ \overbrace{w p^w}^{\text{full accept}}
\end{equation*}
\noindent 
which is a summation of the expected generated tokens when the token is accepted
at different positions.
    

Put it all together, the expectation of TGS is: \\[-5pt]
\begin{equation*}
    \mathrm{TGS}_{g_d,g_v,w}(b) = \frac{\tau_w}{\mathrm{IL}_{g_d,g_v,w}(b)}
\end{equation*}

\begin{algorithm}[t]
\caption{Decoupled execution plan generation algorithm at the start of the rollout. $\mathrm{TGS}_{g_d,g_v,w}(b)$,
$V_{g_v,w}(b)$ and $D_{g_d}(b)$ are descried in \textsection{\ref{sec:design-decouple}}. }
\label{alg:argmax-select}
\KwIn{%
Initial global batch size ${B}$, 
the number of GPUs in the cluster $G$,
a set of execution configurations for verification $\mathbb{G}$.
}
\KwOut{%
    GPUs for drafting $g_{d}^*$, GPUs for verification $g_{v}^*$ and draft window $w^*$.
}


    $\mathrm{TGS}^* \gets 0$, $(g_{d}^*, g_{v}^*, w^*) \gets (0, 0, 0)$\;
    \ForEach{GPU number $g_v \in \mathbb{G}$}{
    \ForEach{GPU number $g_d \gets 1$ \KwTo $g_v$}{
        $b \gets \lceil \frac{(g_d + g_v)B}{G} \rceil$\;
        $w_{max} = \max\{\lceil \frac{V'_{g_v, w}}{D'_{g_d}} \rceil, \lceil \frac{\beta_{g_v, w}}{\alpha_{g_d}} \rceil\}$\;
        \For{$w \gets 1$ \KwTo $w_{\max}$}{
            $\mathrm{TGS}_{\text{cur}} \gets \mathrm{TGS}_{g_d,g_v,w}(b)$\;
            
            \If{$\mathrm{TGS}_{\text{cur}} > \mathrm{TGS}^*$}{
                update $\mathrm{TGS}^*$ and $(g_{d}^*, g_{v}^*, w^*)$\;
            }
        }
    }
    }
\Return{$(g_{d}^*, g_{v}^*, w^*)$}
\end{algorithm}

\begin{algorithm}[t]
\caption{Request-level Reconfiguration for Reducing mis-speculation overheads during the rollout. }
\label{alg:draft-adapt}

\KwIn{%
    pre-searched decoupled plan ($g_d^*, g_v^*, w^*$), \\
    the set of requests with lower acceptance rate than average $\mathbb{R}$.
}

\KwOut{%
    Per-request draft plan $\{(w_r, m_r)\}_{r \in \mathbb{R}}$,\\
    $w_r$ is request draft window, $m_r$ is a flag specifying coupled ($\textsc{c}$) or decoupled ($\textsc{d}$) speculation. 
}


        \ForEach{request $r \in \mathbb{R}$}{
            $p \gets \text{ProfileProbability}(r)$\;
            $w_{\textsc{c}}, \text{tgs}_{\textsc{c}} \gets \displaystyle\arg\max_{w} \text{TGS}_{\textsc{c}, w}(p, b=1)$\;
            $w_{\textsc{d}}, \text{tgs}_{\textsc{d}} \gets \displaystyle\arg\max_{w} \text{TGS}_{g_d^*, g_v^*, w}(p, b=1)$\;
            $w_r, m_r \gets \text{SelectBetter}((w_{\textsc{c}},\text{tgs}_{\textsc{c}}), (w_{\textsc{d}},\text{tgs}_{\textsc{d}}))$
        }

    

\Return{$\{(w_r, m_r): r \in \mathbb{R}\}$}
\end{algorithm}

\stitle{2. Dynamic request-level reconfiguration. \,}
During decoupled execution, we dynamically monitor each request's
acceptance rate to adjust its draft window accordingly. 
Algorithm~\ref{alg:draft-adapt} summarizes our reconfiguration algorithm,
which is called periodically during the rollout.

Upon invocation, the algorithm first examines the set of requests
{with lower acceptance rates than the average (line {1}),}
and reuses the performance model described above for decoupled execution 
and an extended modeling
of coupled execution
 ($\text{TGS}_{\textsc{c}, w}(\cdot)$, omitted due to space limitation,
 since its modeling logic is similar to decoupled modeling)
 to enumerate the best configuration.

Three aspects need to be noted regarding online reconfiguration.
First, we do not trigger reconfiguration too frequently
because (1) the acceptance rate may change sharply over a short period,
and (2) overly frequent reconfiguration may introduce performance instability.
Currently, we reconfigure the system every 1000 decoding iterations.
Second, we execute requests with different window sizes concurrently
to maximize GPU utilization.
To achieve efficient co-execution, we schedule kernels with different
draft windows into a fused CUDA graph~\cite{vllm-code,turbospec}.
Finally, switching between coupled and decoupled execution is straightforward,
as we only need to pause the aggressive draft of the switched request.

\subsection{Effective Fastest-of-N (FoN) Speculative Rollout}
\label{sec:design-bon}

\noindent
This section describes how we select draft methods for the rollout.
The goal is to select the method with the highest speedup.
The challenge lies in balancing the draft efficiency and the acceptance rate,
given that the acceptance rate is unknown prior to execution for a given request.

To tackle this challenge, we leverage the fact that
it is possible to estimate the speedup of different draft methods
parameterized by the acceptance rate
via offline profiling.
With this information, selecting draft methods greedily
using the average acceptance rate of a batch request
is possible at the beginning of the rollout,
and we can further improve the selection
based on the runtime-collected acceptance rate.

\begin{figure}[!t]
        \begin{minipage}{1\linewidth}
        \hspace{-1mm}
        \centering    
        \includegraphics[width=0.85\columnwidth,trim=0.25cm 19.75cm 24cm 0.25cm, clip]{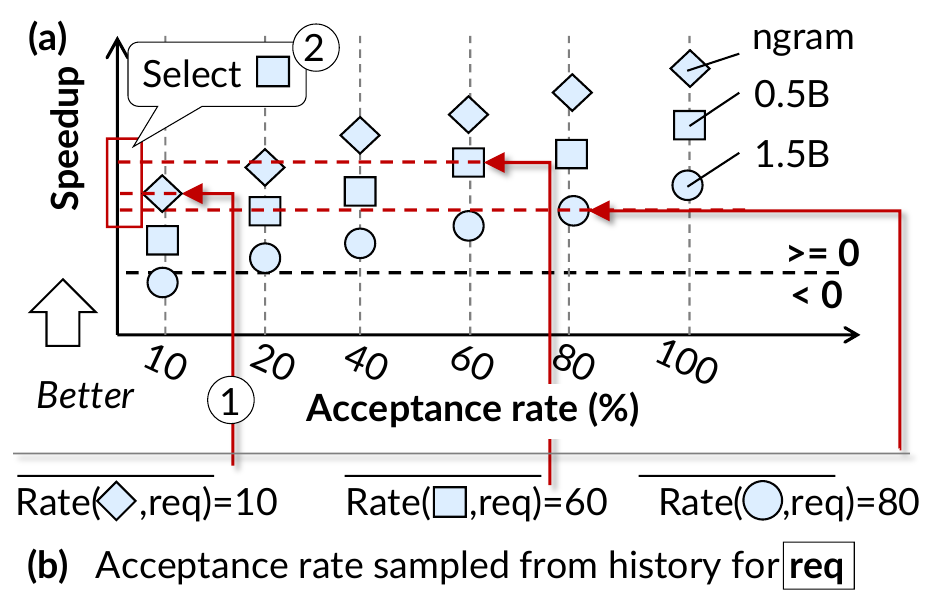} \\[3pt]
        \end{minipage} \\[10pt]
        \begin{minipage}{1\linewidth}
        \caption{\small{%
        {
            (a) Our draft ladder provides hints to the speedup of different draft methods and 
            (b) our ranking (\ding{192}) and selection (\ding{193}) mechanism for choosing
            a draft method. 
        }}}
        \label{fig:draft-ladder}
        \end{minipage} \\[-0pt]
\end{figure}

\stitle{Draft ladder. \,}
A draft ladder provides information on the speedup of different draft methods
given a fixed acceptance rate,
as shown in {\fig{fig:draft-ladder} (a)}.
It can be efficiently constructed via offline profiling
without using the trained model because
(1) the execution of the drafter is irrelevant to the training model and
(2) the speedup can be simulated by randomly accepting tokens according 
to a given acceptance rate.
Thus, our offline profiler directly run the draft methods
with simulated acceptance rate to build the ladder. 

Our current ladder considers popular draft methods, 
including model-based~\cite{spd,specinfer} and n-gram-based~\cite{pld,lookahead,DBLP:conf/acl/HuWZZLCZ25}. 
{For model-based draft, we use a checkpoint from the same series as the large model, 
released jointly before the post-training pipeline.
The practice is similar to previous inference works~\cite{spd,specinfer},
but we cannot use the thinking version draft models as they are typically
distilled after the large thinking model is ready~\cite{deepseek-r1,qwen3}.}
{We have also considered more draft methods,
including training-based drafters
such as the EAGLE family~\cite{eagle,eagle-2,eagle-3}.}
{We do not include MTP~\cite{DBLP:journals/corr/abs-2412-19437} as it is not 
integrated by our evaluated models.}
Nevertheless, we should emphasize that our design
is orthogonal to the draft methods,
and integrating more draft methods is likely to
make {\sys} faster if they can further improve speculation speedup.

\stitle{Draft method selection at the beginning of the rollout. \,}
Initially, we select \emph{one} draft method for
the entire batch based on the historically profiled acceptance rate.
{The profiling only needs to be done once, 
as the profiled results are stable across different steps 
(see \fig{fig:motiv-acc-len}).}
We cannot select multiple methods because each drafter requires 
similar amount of verification, 
and we have insufficient GPU computational power
during the initial phase.

{\fig{fig:draft-ladder} (b)} illustrates the selection process:
For three exemplified draft methods---n-gram, drafting with a 0.5B
and 1.5B model,
we first query the profiled acceptance rate of these methods,
and use these rates as the estimated acceptance rate of the batch.
Based on the rate, we can estimate the speedup of each draft method (\ding{192})
and then select the fastest one among them (\ding{193}).

\begin{algorithm}[t]
\caption{Greedy Fastest-of-N assignments. 
$b_{max}$ is the maximal verification batch size. 
}
\label{alg:bon-sched}

\KwIn{%
    the active requests set $\mathbb{R}$,  
    the candidate draft methods set $\mathbb{D}$.
    $\mathbb{W}_d$ is the set of workers responsible for draft method $d$,
    and $\mathbb{W}$ the whole set of free rollout workers.
}

\KwOut{%
    added drafters and the involved requests $\mathcal{M} = \{(r,d):w\}_{r \in \mathbb{R}, d \in \mathbb{D}, w \in \mathbb{W}}$.
}

$\mathcal{R} \gets$ sort $\mathbb{R}$ by \text{GetAcceptRate}(r) ascending\;
$\mathcal{D} \gets$ sort $\mathbb{D}$ by \text{GetLadderRank}(d) ascending\;
\ForEach{request $r\in\mathcal{R}$}{
\ForEach{draft method $d\in\mathcal{D}$}{
    \If{$\mathcal{M}(r,d)$ is None}{
            $w \gets \text{GetMinLoadWorker}(\mathbb{W}_d, b_{max})$\;
            \If{$w$ is not None}{
                $\mathcal{M}(r,d) \gets w$\;
                $\text{load}(w)\mathrel{+}=1$\;
            }
    }
}
}

\Return{$\mathcal{M}$}
\end{algorithm}

\stitle{Greedy Fastest-of-N assignments. \,}
To cope with the sub-optimal initial selection for tailed requests,
we greedily deploy more draft methods as soon as
GPUs of both drafter and verifier become available upon batch completion.
\alg{alg:bon-sched} shows our algorithm for assigning
more drafters (and their corresponding verifier) when there are available workers.
The routine is iteratively called once there are free workers.

The algorithm assigns drafters to requests in an order sorted
by the reverse of their acceptance rate (lines 1)
as these requests have little speedup with speculative decoding.
For each request, we add a drafter with the highest
speedup according to the draft ladder described before 
if it has not been assigned yet (line 2 and 5).
We greedily assign as many drafters as possible to the request
until there are no available workers or pending requests (lines 3--9).
If the assignments for a request are completed,
we proceed to the next request.

Here we adopt a draft-first design that assigns as many draft methods to one request as possible before moving on to the next one.
The rationale is that the request with the lowest acceptance rate is likely to suffer from suboptimal drafting,
so the draft-first strategy can naturally select the optimal draft method for it (given a sufficient number of free workers).
In case of an insufficient number of free workers, we prioritize the draft method with a higher rank in our draft ladder.

\stitle{{Discussion for evolvable draft methods.} \,}
Our ladder assumes that the draft models are frozen during the post-training, 
because we found that the acceptance rate of draft methods remains stable 
across steps (even with long gap), see {\fig{fig:motiv-acc-len}}. 
We suspect this is because even as the trained model becomes smarter,
many trivial tokens can still be drafted as long as the trained model corrects
the incorrect tokens timely. 
{Note that recent draft training optimizations~\cite{tlt} is orthogonal 
to our design as they can also be applied to our small models.}
Further, integrating such evolving would require
intrusive modifications to both the rollout and training engines.
Therefore, we don't include them in our implementation. 

\subsection{Efficient System Runtime for {\sys}}
\label{sec:design-mechanisms}

\noindent 
{\sys} only optimizes the rollout phase and makes no intrusive modifications
to the training engine, in contrast to~\cite{rlhfuse,rollpacker}.
The runtime extension is also lightweight:
supporting decoupled execution only requires message passing
between the drafter and the verifier,
and we further incorporate two primitives to support Fastest-of-N speculation:

\stitle{Model scale. \,}
The primitive deploys a serving instance—--either a drafter or a verifier—to a specific worker.
This is similar to model autoscaling in existing serving engines~\cite{DBLP:conf/osdi/FuXHBUPM24,DBLP:conf/sosp/Xiang0QYZYZL0025,DBLP:conf/sosp/WeiHSH0H0025,DBLP:conf/asplos/ZengXGCL25,DBLP:conf/osdi/ZhangWLWS0025} 
and we adopted all their designs, 
including pre-pinning serving engines on workers~\cite{DBLP:conf/sosp/Xiang0QYZYZL0025},
pre-materializing CUDA graphs with GPU execution context pools~\cite{DBLP:conf/asplos/ZengXGCL25,DBLP:conf/sosp/WeiHSH0H0025}
and transfer model weights via fast networking between GPUs~\cite{DBLP:conf/osdi/ZhangWLWS0025}. 
To further eliminate the overhead of loading large trained model weights to GPUs
especially in case of deploying a new verifier, 
we observed that verifiers are deployed only to freed drafters and the 
drafter GPUs have low GPU memory usage due to the small size of draft models. 
Thus, we further pin the trained models to drafters to enable a zero-cost verifier deployment.

\stitle{KVCache scale. \,}
Besides the model, when scaling a new drafter's corresponding verifier,
the new verifier requires a copy of the KVCache---the intermediate results
of the LLM computation---to accelerate computation.
A challenge is that during the later stage of rollout,
the KVCache may become large since its size is proportional to the model size and the number of
generated tokens.
To this end, we can leverage the classic KVCache recovery method,
which transfers the tail KVCache through the network
and recomputes it from the beginning~\cite{DBLP:journals/corr/abs-2410-03065}
to accelerate the KVCache scaling.

\section{Evaluation}
\label{sec:eval}

\subsection{Experimental setup}
\label{sec:eval-setup}

\noindent
We implemented {\sys} on {\verl}---the state-of-the-art
post-training framework that
has integrated known inference optimizations
including CUDA graph~\cite{vllm-code},
prefix caching~\cite{DBLP:journals/corr/abs-2402-03300} 
and FlashAttention kernels~\cite{DBLP:conf/nips/ShahBZTRD24}. 

\stitle{Testbed. \,} 
We evaluated {\sys} on a production training cluster
with up to 512\,GPUs.
These GPUs are organized into nodes where each node contains eight Hopper (80\,GB) GPUs
interconnected with {400}\,GB/s NVLink,
and inter-node GPUs have up to 400\,Gbps RDMA connections.

\stitle{Evaluated traces: model, algorithm and rollout temperature.} \,
Similar to prior works~\cite{dapo,hybridflow,streamrl}, 
we evaluated Qwen series models (\eg{Qwen2.5-32B})
as they are one of the most popular open-sourced LLMs~\cite{dapo,DBLP:journals/corr/abs-2504-05118,gspo}
and are commonly used by algorithm designers. 
We evaluated on {3} 
representative 200-step training traces based on production setups
with different (popular) post-training algorithms: \\[-18pt]

\begin{itemize}[leftmargin=*,leftmargin=10pt,itemindent=0pt]
    \itemsep0.5em
    \item \textbf{GRPO-32B-20K.} 
    This trace trains Qwen2.5-32B~\cite{qwen2.5}
    with the GRPO~\cite{DBLP:journals/corr/abs-2402-03300} algorithm,
    a representative value-model-free post-training method 
    that trains high-quality models like DeepSeek series~\cite{DBLP:journals/corr/abs-2509-02547,deepseek-r1}.
    The trace samples a total batch of 8,192 prompts at each step\footnote{\footnotesize{Including the group sampling factor. }},
    where for each batch, the model generates responses with a budget of 20\,K tokens.
    The trace runs on 256\,GPUs with a tensor parallelism (TP) degree of 4 for
    each rollout worker, so the initial per-worker batch size is 128.  \\[-18pt]

    \item \textbf{DAPO-32B-20K.} 
    This trace also trains Qwen2.5-32B using DAPO~\cite{dapo}---a
    trending algorithm in AI academy and industry~\cite{DBLP:journals/corr/abs-2508-03501,DBLP:journals/corr/abs-2505-23433}---with
    a per-step batch size of 16,384 and a response budget of 20\,K tokens.
    The trace runs on 256 GPUs with a 4-degree TP for each rollout worker,
    so the per-worker batch size is 256.
    DAPO requires a larger per-step batch size because its training
    method filters out low-quality responses generated during the rollout. \\[-18pt]

    \item \textbf{PPO-32B-20K.} 
    PPO is another widely used RL algorithm~\cite{DBLP:journals/corr/SchulmanWDRK17,DBLP:journals/corr/abs-2504-05118}.
    The trained model is Qwen2.5-32B, and the PPO trace is different in 
    (1) it utilizes another 32B critic model to generate value and trains the critic together with the actor model;
    (2) meanwhile, it samples only one response per prompt. The per-step batch size is 4,096 
    and the maximum response length is 20\,K tokens. 
    We run the trace on the same cluster setup like the above two traces. \\[-14pt]
\end{itemize}

\noindent
For all experiments, we set the sampling temperature of LLM generation
to 1.0---a common setup in post-training~\cite{deepseek-r1,DBLP:journals/corr/abs-2402-03300,DBLP:journals/corr/abs-2507-20534}---which, however, 
has a negative impact on speculative decoding
~\cite{DBLP:conf/acl/XiaYDW00L0S24,DBLP:journals/corr/abs-2509-04474,DBLP:journals/corr/abs-2510-26475}.
LLM post-training requires such settings to explore more diverse responses
and avoid entropy collapse~\cite{DBLP:journals/corr/abs-2505-23585,dapo,DBLP:journals/corr/abs-2504-05118};
otherwise, the trained model suffers from premature convergence, 
which limits its potential for further performance gains.

Due to GPU time constraints, our experiments focus on dense 32\,B
models, but we have also conducted experiments on a large 
Qwen3-235B MoE model~\cite{qwen3}
in \textsection{\ref{sec:eval-moe}} to demonstrate the effectiveness of {\sys}. 

\stitle{Evaluating metrics. \,}
We focus on reporting the end-to-end training time and the rollout time---since
training time is the most critical metric for algorithm developers
and rollout dominates the time in \textsection{\ref{sec:bg-post-training}}.
To ensure representative evaluation, we uniformly sample at least 10\% of all training steps,
using the trained checkpoints at the sampled steps for generation,
and report their training and rollout time.

\stitle{Baselines. \,}
We compared {\sys} with the following baselines and carefully tuned their
implementations and performance.
All systems rollout with vLLM v0.10.0 engine~\cite{vllm-code}.
Specifically,
the decoding latency of Qwen2.5-32B model
is lower to 13\,ms when per-worker batch size is 1 on our platform. 
For all baselines, we report the optimal configuration for training,
i.e., an FSDP degree of 32 and 8-way sequence parallelism~\cite{ulysses-sp}.
\\[-10pt]

\begin{itemize}[leftmargin=*,leftmargin=10pt,itemindent=0pt]
    \itemsep0.5em
    \item \textbf{\verl~\cite{DBLP:conf/eurosys/ShengZYWZZPL025}}
    is an open-sourced post-training system used by {\company}
    and {\sys} is based on it.
    Besides the inference optimizations mentioned in the beginning of the section,
    it further incorporates techniques like efficient parameter
    to quickly update parameters for the rollout. \\[-10pt]

    \item \textbf{RLHFuse~\cite{rlhfuse}} 
    is the state-of-the-art post-training system that overlaps
    the prepare and rollout phases to reduce bubbles caused by
    long-tailed generation.
    Since it is not open-sourced, we carefully re-implemented its design
    and achieve a similar speedup on short-context tasks. 
    \\[-18pt]

    \item \textbf{{\verl} (2\,$\times$)} 
    provisions 2\,$\times$ of GPUs used by the original trace for {\verl}, 
    which serves as the optimal performance of using more GPUs 
    to accelerate RL like RLBoost~\cite{rlboost}.  \\[-18pt]

    \item \textbf{{\verl} + (vanilla) model-spec.}
    To illustrate the effectiveness of our proposed techniques,
    we also compare {\verl} by incorporating model-based speculative
    decoding in inference systems~\cite{spd,specinfer}.
    We run coupled speculation described in \textsection{\ref{sec:overview}}.
    The draft method is selected based on the first phase of our Fastest-of-N speculation.
    For 32B training, we found 0.5B is a sweet point to accelerate. \\[-18pt]

    \item \textbf{{\verl} + (vanilla) n-gram.}
    N-gram is a popular speclative decoding method used by other works~\cite{rhymerl,seer}
    to accelerate rollout. 
    We have integrated a baseline of it using vLLM's n-gram (v0.10.0), 
    with enhancement of recent techniques like SAM~\cite{DBLP:conf/acl/HuWZZLCZ25} 
    to match their implementation as much as possible (since their code is not open-sourced). \\[-18pt]

\end{itemize}

\stitle{Draft methods used by {\sys}. \,}
Unless otherwise specified, 
we select the available drafters for Qwen2.5-32B 
before the post-training pipeline:
Qwen2.5-0.5B~\cite{qwen2.5}, Qwen2.5-1.5B---as well as an n-gram drafter~\cite{vllm-code,DBLP:conf/acl/HuWZZLCZ25}.
Other drafting methods, such as EAGLE~\cite{eagle,eagle-2,eagle-3}, 
require additional training support,
which is not supported by all our evaluated models. 


\subsection{End-to-end post-training performance}
\label{sec:eval-e2e}

\begin{figure}[!t]
        \begin{minipage}{1\linewidth}
        \hspace{-1mm}
        \centering    
        \includegraphics[width=\columnwidth]{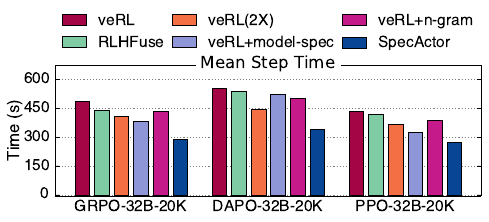} \\[1pt]    
        \end{minipage} \\[0pt]
        \begin{minipage}{1\linewidth}
        \caption{\small{%
        {
            The mean training step time
            of {\sys} running different training traces
            with different approaches. 
        }}}
        \label{fig:e2e}
        \end{minipage} \\[-15pt]
\end{figure}

\begin{figure}[!t]
        \begin{minipage}{1\linewidth}
        \hspace{-1mm}
        \centering    
        \includegraphics[width=\columnwidth]{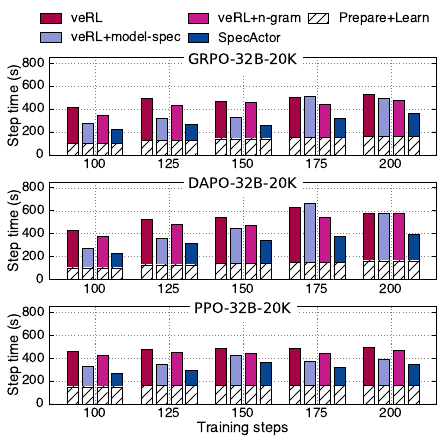} \\[1pt]    
        \end{minipage} \\[1pt]
        \begin{minipage}{1\linewidth}
        \caption{\small{%
        {
        A breakdown of post-training process of different approaches on
        different training traces.
        }}}
        \label{fig:breakdown}
        \end{minipage} \\[-15pt]
\end{figure}

\noindent
\fig{fig:e2e} presents the end-to-end training step time of different approaches
on the traces. 
We measure all 1/10 of the sampled training steps and report the mean step time of 
different approaches. 
Note that each step uses the exact same model checkpoint and prompt batch 
in the trace to faithfully reproduce the real-world workloads. 
{\sys} achieves {1.5--1.7\,$\times$} faster end-to-end 
training compared to baselines like veRL and RLHFuse.
For vanilla speculative decoding baselines, 
{\sys} achieves {1.2--1.5\,$\times$} shorter training time in different traces.
Even compared with {\verl} (2\,$\times$) that uses twice the GPUs, 
{\sys} still achieves {1.3--1.4\,$\times$} training speedup.
The key contributor is the acceleration of rollout via our
efficient mechanisms: 
for the rollout phase, {\sys} achieves {2.0--2.4}\,$\times$ faster 
mean generation speed than {\verl}.

Notably, {\sys} achieves up to {1.4--2.0\,$\times$} shorter average rollout time 
even compared with methods that also incorporate speculative decoding acceleration,
\ie{model-spec and n-gram}.
This is because:
First, both of them suffer from inefficiency due to large per-worker batch size
in the initial phase.
Second, both of them, especially n-gram, have low acceptance rates for the straggler requests,
because the n-gram method performs poorly in high temperature sampling
with few history prompts commonly found in training traces.
Specifically, the skipped iteration of the last finished requests of n-gram 
is {16.9--43.6}\,\% across the traces,
{while {\sys} achieves 40.9--73.5\,\%
thanks to our Fastest-of-N speculation.}

\begin{figure}[!t]
       \vspace{2mm}
        \begin{minipage}{1\linewidth}
        \hspace{-1mm}
        \centering    
        \includegraphics[width=\columnwidth]{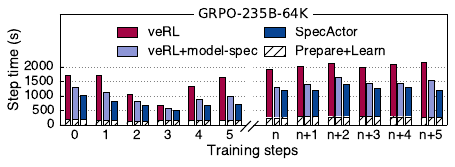} \\[1pt]    
        \end{minipage} \\[1pt]
        \begin{minipage}{1\linewidth}
        \caption{\small{%
        {
        A breakdown of post-training steps of Qwen3-235B.
        }}}
        \label{fig:moe}
        \end{minipage} \\[-15pt]
\end{figure}

\subsection{Performance on large MoE models}
\label{sec:eval-moe}

\noindent
\fig{fig:moe} presents the performance of {\sys}
when training a Qwen3-235B MoE model~\cite{qwen3} on 256\,GPUs
with GRPO algorithm.
Due to limited GPU time and the availability of checkpoints,
we only evaluated a few steps:
0--5 at the start of the post-training (using the Qwen3-235B-Instruct-2507)
and $n, \ldots, n+5$ from the end of the training (using the Qwen3-235B-Thinking-2507).
The difference is that the later steps tend to generate longer responses.
The per-step batch size is 256---a slightly smaller batch size due to
limited GPU memory, and each rollout worker uses expert parallelism (EP) of 8
for high decode efficiency.

{We expand the draft ladder with small models produced before
post-training pipeline~\cite{qwen3}: Qwen3-4B-2507-Instruct 
(released together with 235B), Qwen3-0.6B and Qwen3-1.7B.}

{\sys} improves the training time by {1.4--2.3}\,$\times$,
with {1.5--2.6}\,$\times$ rollout acceleration compared to {\verl}.
It also outperforms vanilla mode-spec by {1.1--1.5}\,$\times$, 
thanks to our decoupled and Fastest-of-N speculation.
Notice that even with a relatively small per-step batch size,
verification overhead is still high in MoE models as it is exacerbated
by expert communication~\cite{DBLP:journals/corr/abs-2505-19645}. 
{We also found that 
Qwen3-4B-2507-Instruct performs significantly better in speculation 
than the other two, we suspect its training pipeline is closely coupled
with 235B and thus has better alignment with it.}

\subsection{Performance across different steps}
\label{sec:eval-breakdown}

\noindent
As the trained model evolves across different training steps,
the effectiveness of speculative decoding varies.
Thus, we further present the detailed latency breakdown of different
training steps in \fig{fig:breakdown}.
For ease of presentation, we selectively
present part of the training steps between {100--200} training steps to examine speculative decoding's effect
in real-world post-training.
The gap is sufficiently large as confirmed by internal accuracy tests,
so the "smartness" of the trained models in different sampling
steps varies.

First we can see that 
{\sys} still achieves the shortest rollout time in all steps.
Specifically, at $175^{th}$ and $200^{th}$ steps, 
veRL + model-spec and n-gram 
fail to provide acceleration because 
as models becomes smarter, they tend to produce longer responses~\cite{deepseek-r1} for more prompts,
leading to more rollout steps under a relatively large per-worker batch size.
In contrast, {\sys} consistently reduces rollout time by {1.8--2.7}\,$\times$,
and is faster than vanilla speculative decoding baselines by 
{1.4--2.6}\,$\times$, {1.2--2.2}\,$\times$, {1.2--2.6}\,$\times$ in different training traces.
Thanks to the quick generation,
more workers finish at early time,
leaving {\sys} sufficient room for FoN scheduling. 

\subsection{Ablation study}

\begin{figure}[!t]
        \begin{minipage}{1\linewidth}
        \hspace{-1mm}
        \centering    
        \includegraphics[width=\columnwidth]{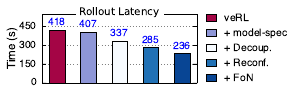} \\[1pt]    
        \end{minipage} \\[0pt]
        \begin{minipage}{1\linewidth}
        \caption{\small{%
        {            
            Ablation study of using the DAPO-32B-20K trace.
            The other trace shows similar results.
        }}}
        \label{fig:ablation}
        \end{minipage} \\[-15pt]
\end{figure}

\noindent
\fig{fig:ablation} conducts an ablation study to
examine the effectiveness of different proposed techniques described in \textsection{\ref{sec:design}}.
We reported only one step here due to space limitation, and other steps in different traces 
show similar results.

First, we can see that vanilla speculative decoding reduces only {2.6\,\%} rollout end-to-end latency,
which is small due to the inefficiency of large-batch verification 
and suboptimal draft method for the tailed requests. 

Thanks to our decoupled speculative decoding,
we accelerate rollout by {1.3\,$\times$}.
Based on decoupled speculative decoding,
our dynamic reconfiguration
further shortens the rollout time by {1.2\,$\times$} thanks
to a better initial execution configuration and dynamic reconfiguration
of workers with low acceptance rates to reduce wasted tokens.
Finally, our Fastest-of-N speculative decoding further accelerates end-to-end rollout
by {1.2\,$\times$}, thanks to its ability to leverage multiple draft methods
to improve acceptance rates for straggler requests.

\begin{figure}[!t]
        \begin{minipage}{1\linewidth}
        \hspace{-1mm}
        \centering    
        \includegraphics[width=\columnwidth]{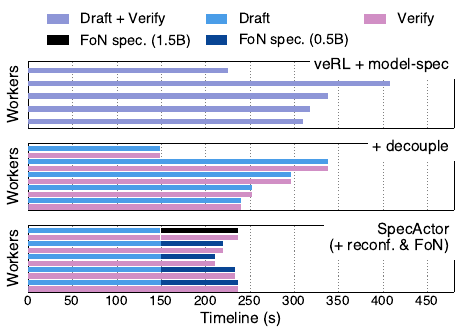} \\[1pt]    
        \end{minipage} \\[1pt]
        \begin{minipage}{1\linewidth}
        \caption{\small{%
        {
    An in-depth analysis of the execution of different approaches on
    the {$200^{th}$}  step of DAPO-32B-20K trace. 
    We sample workers that contains the tailed requests for ease of presentation.
        }}}
        \label{fig:timeline}
        \end{minipage} \\[-5pt]
\end{figure}

\subsection{An in-depth look at {\sys} in action}
\label{sec:eval-timeline}

\noindent
We conclude our evaluation of {\sys} by presenting an in-depth analysis
of the work done by different workers during the rollout with our techniques in
\fig{fig:timeline}.
To ease presentation, we only sampled {5} workers,
with a deliberate focus on the worker that finish early
and the slowest four workers.

First, from the first row of \fig{fig:timeline}
we can see that vanilla speculative decoding still
suffers from long-tailed workers
because during the initial phase all workers execute slower
and at the end the initially selected draft method
cannot accelerate the slowest requests.
With decoupled speculative decoding (the second row), we improve speculative decoding performance
by {1.1--1.5}\,$\times$ but still suffer from long-tailed requests
due to suboptimal draft method.
With Fastest-of-N speculative decoding activated, the freed workers are assigned
different draft methods (we use different color blocks to indicate different draft methods)
at 151s in \fig{fig:timeline}.
This subsequently accelerates the slowest requests by {1.5}\,$\times$.

\section{Related work}
\label{sec:related-work}

\nospacestitle{Rollout acceleration with algorithmic modifications. \,}
Several works propose algorithmic changes to accelerate rollout,
e.g., skipping the tail generations~\cite{DBLP:journals/corr/abs-2501-12599,DBLP:journals/corr/abs-2504-13914,rhymerl}.
Such modifications essentially make the training 
a variant of off-policy (vs. on-policy of our targeted setup),
which has unstable convergence in theory~\cite{DBLP:journals/corr/abs-2506-20520,DBLP:journals/corr/abs-2503-14286,DBLP:journals/corr/abs-2405-08448}
and {we observe that they are not always adopted in production~\cite{seer,deepseek-r1}.}

\stitle{Rollout acceleration for resource-restrict setups. \,}
Recent works like Seer~\cite{seer} 
and RollPacker~\cite{rollpacker} address long-tail generation 
in resource-constraint setups,
i.e., each worker has limited GPU capacity to hold the 
assigned batch. 
As a result, the batch must be further divided into sub-batches where 
each sub-batch is executed sequentially,
and their optimizations lie in reordering or repacking
sub-patches to reduce GPU wastes between sub-batch executions. 
In contrast, {\sys} accelerates long-tail generations where the batch can
fit in the GPU memory of each worker, a common production setup (see {\fig{fig:motiv-batchsize}} (a)).
In general, executing one large batch is faster than sequentially
executing the divided sub-batch (if with sufficient resources) 
because the generation time grows sub-linearly with the batch size 
(see {\fig{fig:motiv-batchsize}} (b)), 
not to mention the possible switching overhead between sub-batches 
with packing techniques~\cite{rollpacker}.

\stitle{Concurrent works on accelerating rollout via speculative decoding. \,}
A few concurrent works 
share a similar high-level idea of using speculative decoding to accelerate rollout~\cite{tlt,seer}.
{\sys} further addresses the low efficiency issue in batch configurations that are unfriendly
to speculative decoding,
and the unknown best drafter selection issue across requests.

The closest work to ours is TLT~\cite{tlt}, 
which selects training-based drafter (EAGLE~\cite{eagle,eagle-2,eagle-3}) 
for speculative rollout. 
It does not consider the large batch efficiency issue and 
its performance can be further improved with our decoupled speculation.
Moreover, our draft ladder with Fastest-of-N speculation 
further provides two advantages over its chosen EAGLE:
(1) EAGLE is more challenging to deploy than the model- and n-gram-based methods 
in our ladder because it requires extra training to match the target model~\cite{eagle-2,eagle-3}, 
which is not always available for the trained model or requires extra tuning; and
(2) we found it has a lower acceptance length than the drafters 
we considered using the models provided by TLT without prompt tuning for EAGLE~\cite{tlt-code} (see \fig{fig:motiv-acc-len}).
Finally, TLT adopts approximate sampling to improve the acceptance rate 
at the cost of training accuracy~\cite{DBLP:journals/corr/abs-2510-26475}.
In contrast, {\sys} uses exact matching to ensure the correctness of the rollout responses.

\stitle{Speculative decoding in inference. \,}
The optimization goal of using speculative decoding 
in rollout is fundamentally different from that in inference systems
~\cite{spd,specinfer,pearl,swiftspec,DBLP:conf/icml/LiuHBCDS024,turbospec}. 
Specifically, rollout prioritizes overall batch execution efficiency, 
where individual request latencies are less critical, 
while inference systems must ensure low latencies for all requests. 
This introduces new challenges, which we address with decoupled and 
Fastest-of-N speculation.

\section{Conclusion}
\label{sec:concl}

\noindent
This paper shows that RL rollout can be efficient 
without compromising the algorithmic equivalence of the original training with 
fast speculative decoding. 
{\sys} makes speculative rollout efficient with two novel techniques: 
(1) a \emph{Decoupled speculation} method that improves the speculation efficiency 
for common training setups, 
and (2) a \emph{Fastest-of-N speculation} method that 
dynamically adapts the draft method
during the rollout process.
{\sys} improves the training efficiency by {1.4--2.3}\,$\times$ compared to the state-of-the-art baselines,
and is {1.1--2.6}\,$\times$ faster than vanilla speculative rollout.

\balance

\small{
\bibliographystyle{acm}
\bibliography{specrl}

@article{hoeffding1963probability,
  title={Probability inequalities for sums of bounded random variables},
  author={Hoeffding, Wassily},
  journal={Journal of the American statistical association},
  volume={58},
  number={301},
  pages={13--30},
  year={1963},
  publisher={Taylor \& Francis}
}

@inproceedings{DBLP:conf/acl/XiaYDW00L0S24,
  author       = {Heming Xia and
                  Zhe Yang and
                  Qingxiu Dong and
                  Peiyi Wang and
                  Yongqi Li and
                  Tao Ge and
                  Tianyu Liu and
                  Wenjie Li and
                  Zhifang Sui},
  editor       = {Lun{-}Wei Ku and
                  Andre Martins and
                  Vivek Srikumar},
  title        = {Unlocking Efficiency in Large Language Model Inference: {A} Comprehensive
                  Survey of Speculative Decoding},
  booktitle    = {Findings of the Association for Computational Linguistics, {ACL} 2024,
                  Bangkok, Thailand and virtual meeting, August 11-16, 2024},
  pages        = {7655--7671},
  publisher    = {Association for Computational Linguistics},
  year         = {2024},
  url          = {https://doi.org/10.18653/v1/2024.findings-acl.456},
  doi          = {10.18653/V1/2024.FINDINGS-ACL.456},
  bibsource    = {dblp computer science bibliography, https://dblp.org}
}

@article{DBLP:journals/corr/abs-2410-03065,
  author       = {Shuowei Jin and
                  Xueshen Liu and
                  Qingzhao Zhang and
                  Z. Morley Mao},
  title        = {Compute Or Load {KV} Cache? Why Not Both?},
  journal      = {CoRR},
  volume       = {abs/2410.03065},
  year         = {2024},
  url          = {https://doi.org/10.48550/arXiv.2410.03065},
  doi          = {10.48550/ARXIV.2410.03065},
  eprinttype    = {arXiv},
  eprint       = {2410.03065},
  bibsource    = {dblp computer science bibliography, https://dblp.org}
}

@inproceedings{DBLP:conf/osdi/ZhuGZZZGX0KLWWK25,
  author       = {Kan Zhu and
                  Yufei Gao and
                  Yilong Zhao and
                  Liangyu Zhao and
                  Gefei Zuo and
                  Yile Gu and
                  Dedong Xie and
                  Zihao Ye and
                  Keisuke Kamahori and
                  Chien{-}Yu Lin and
                  Ziren Wang and
                  Stephanie Wang and
                  Arvind Krishnamurthy and
                  Baris Kasikci},
  editor       = {Lidong Zhou and
                  Yuanyuan Zhou},
  title        = {NanoFlow: Towards Optimal Large Language Model Serving Throughput},
  booktitle    = {19th {USENIX} Symposium on Operating Systems Design and Implementation,
                  {OSDI} 2025, Boston, MA, USA, July 7-9, 2025},
  pages        = {749--765},
  publisher    = {{USENIX} Association},
  year         = {2025},
  url          = {https://www.usenix.org/conference/osdi25/presentation/zhu-kan},
  bibsource    = {dblp computer science bibliography, https://dblp.org}
}

@inproceedings{DBLP:conf/sosp/Xiang0QYZYZL0025,
  author       = {Yuxing Xiang and
                  Xue Li and
                  Kun Qian and
                  Yufan Yang and
                  Diwen Zhu and
                  Wenyuan Yu and
                  Ennan Zhai and
                  Xuanzhe Liu and
                  Xin Jin and
                  Jingren Zhou},
  editor       = {Youjip Won and
                  Youngjin Kwon and
                  Ding Yuan and
                  Rebecca Isaacs},
  title        = {Aegaeon: Effective {GPU} Pooling for Concurrent {LLM} Serving on the
                  Market},
  booktitle    = {Proceedings of the {ACM} {SIGOPS} 31st Symposium on Operating Systems
                  Principles, {SOSP} 2025, Lotte Hotel World, Seoul, Republic of Korea,
                  October 13-16, 2025},
  pages        = {1030--1045},
  publisher    = {{ACM}},
  year         = {2025},
  url          = {https://doi.org/10.1145/3731569.3764815},
  doi          = {10.1145/3731569.3764815},
  bibsource    = {dblp computer science bibliography, https://dblp.org}
}

@inproceedings{DBLP:conf/osdi/ZhangWLWS0025,
  author       = {Dingyan Zhang and
                  Haotian Wang and
                  Yang Liu and
                  Xingda Wei and
                  Yizhou Shan and
                  Rong Chen and
                  Haibo Chen},
  editor       = {Lidong Zhou and
                  Yuanyuan Zhou},
  title        = {BlitzScale: Fast and Live Large Model Autoscaling with {O(1)} Host
                  Caching},
  booktitle    = {19th {USENIX} Symposium on Operating Systems Design and Implementation,
                  {OSDI} 2025, Boston, MA, USA, July 7-9, 2025},
  pages        = {275--293},
  publisher    = {{USENIX} Association},
  year         = {2025},
  url          = {https://www.usenix.org/conference/osdi25/presentation/zhang-dingyan},
  bibsource    = {dblp computer science bibliography, https://dblp.org}
}

@inproceedings{DBLP:conf/sosp/WeiHSH0H0025,
  author       = {Xingda Wei and
                  Zhuobin Huang and
                  Tianle Sun and
                  Yingyi Hao and
                  Rong Chen and
                  Mingcong Han and
                  Jinyu Gu and
                  Haibo Chen},
  editor       = {Youjip Won and
                  Youngjin Kwon and
                  Ding Yuan and
                  Rebecca Isaacs},
  title        = {PhoenixOS: Concurrent OS-level {GPU} Checkpoint and Restore with Validated
                  Speculation},
  booktitle    = {Proceedings of the {ACM} {SIGOPS} 31st Symposium on Operating Systems
                  Principles, {SOSP} 2025, Lotte Hotel World, Seoul, Republic of Korea,
                  October 13-16, 2025},
  pages        = {996--1013},
  publisher    = {{ACM}},
  year         = {2025},
  url          = {https://doi.org/10.1145/3731569.3764813},
  doi          = {10.1145/3731569.3764813},
  bibsource    = {dblp computer science bibliography, https://dblp.org}
}

@inproceedings{DBLP:conf/osdi/FuXHBUPM24,
  author    = {Yao Fu and
               Leyang Xue and
               Yeqi Huang and
               Andrei{-}Octavian Brabete and
               Dmitrii Ustiugov and
               Yuvraj Patel and
               Luo Mai},
  editor    = {Ada Gavrilovska and
               Douglas B. Terry},
  title     = {ServerlessLLM: Low-Latency Serverless Inference for Large Language
               Models},
  booktitle = {18th {USENIX} Symposium on Operating Systems Design and Implementation,
               {OSDI} 2024, Santa Clara, CA, USA, July 10-12, 2024},
  pages     = {135--153},
  publisher = {{USENIX} Association},
  year      = {2024},
  url       = {https://www.usenix.org/conference/osdi24/presentation/fu},
  bibsource = {dblp computer science bibliography, https://dblp.org}
}

@misc{vllm-code,
  title        = {Easy, fast, and cheap LLM serving for everyone},
  howpublished = {\burl{https://github.com/vllm-project/vllm}},
  year         = {2024}
}

@article{DBLP:journals/corr/SchulmanWDRK17,
  author       = {John Schulman and
                  Filip Wolski and
                  Prafulla Dhariwal and
                  Alec Radford and
                  Oleg Klimov},
  title        = {Proximal Policy Optimization Algorithms},
  journal      = {CoRR},
  volume       = {abs/1707.06347},
  year         = {2017},
  url          = {http://arxiv.org/abs/1707.06347},
  eprinttype    = {arXiv},
  eprint       = {1707.06347},
  bibsource    = {dblp computer science bibliography, https://dblp.org}
}

@inproceedings{DBLP:conf/asplos/ZengXGCL25,
  author       = {Shaoxun Zeng and
                  Minhui Xie and
                  Shiwei Gao and
                  Youmin Chen and
                  Youyou Lu},
  editor       = {Lieven Eeckhout and
                  Georgios Smaragdakis and
                  Kaitai Liang and
                  Adrian Sampson and
                  Martha A. Kim and
                  Christopher J. Rossbach},
  title        = {Medusa: Accelerating Serverless {LLM} Inference with Materialization},
  booktitle    = {Proceedings of the 30th {ACM} International Conference on Architectural
                  Support for Programming Languages and Operating Systems, Volume 1,
                  {ASPLOS} 2025, Rotterdam, The Netherlands, 30 March 2025 - 3 April
                  2025},
  pages        = {653--668},
  publisher    = {{ACM}},
  year         = {2025},
  url          = {https://doi.org/10.1145/3669940.3707285},
  doi          = {10.1145/3669940.3707285},
  bibsource    = {dblp computer science bibliography, https://dblp.org}
}

@misc{qwen2.5,
  title  = {Qwen2.5: A Party of Foundation Models},
  url    = {https://qwenlm.github.io/blog/qwen2.5/},
  author = {Qwen Team},
  month  = {September},
  year   = {2024}
}

@inproceedings{DBLP:conf/nips/SternSU18,
  author       = {Mitchell Stern and
                  Noam Shazeer and
                  Jakob Uszkoreit},
  editor       = {Samy Bengio and
                  Hanna M. Wallach and
                  Hugo Larochelle and
                  Kristen Grauman and
                  Nicol{\`{o}} Cesa{-}Bianchi and
                  Roman Garnett},
  title        = {Blockwise Parallel Decoding for Deep Autoregressive Models},
  booktitle    = {Advances in Neural Information Processing Systems 31: Annual Conference
                  on Neural Information Processing Systems 2018, NeurIPS 2018, December
                  3-8, 2018, Montr{\'{e}}al, Canada},
  pages        = {10107--10116},
  year         = {2018},
  url          = {https://proceedings.neurips.cc/paper/2018/hash/c4127b9194fe8562c64dc0f5bf2c93bc-Abstract.html},
  bibsource    = {dblp computer science bibliography, https://dblp.org}
}

@inproceedings{DBLP:conf/asplos/MiaoOZCWZWZYSSC24,
  author       = {Xupeng Miao and
                  Gabriele Oliaro and
                  Zhihao Zhang and
                  Xinhao Cheng and
                  Zeyu Wang and
                  Zhengxin Zhang and
                  Rae Ying Yee Wong and
                  Alan Zhu and
                  Lijie Yang and
                  Xiaoxiang Shi and
                  Chunan Shi and
                  Zhuoming Chen and
                  Daiyaan Arfeen and
                  Reyna Abhyankar and
                  Zhihao Jia},
  editor       = {Rajiv Gupta and
                  Nael B. Abu{-}Ghazaleh and
                  Madan Musuvathi and
                  Dan Tsafrir},
  title        = {SpecInfer: Accelerating Large Language Model Serving with Tree-based
                  Speculative Inference and Verification},
  booktitle    = {Proceedings of the 29th {ACM} International Conference on Architectural
                  Support for Programming Languages and Operating Systems, Volume 3,
                  {ASPLOS} 2024, La Jolla, CA, USA, 27 April 2024- 1 May 2024},
  pages        = {932--949},
  publisher    = {{ACM}},
  year         = {2024},
  url          = {https://doi.org/10.1145/3620666.3651335},
  doi          = {10.1145/3620666.3651335},
  bibsource    = {dblp computer science bibliography, https://dblp.org}
}

@inproceedings{DBLP:conf/icml/LeviathanKM23,
  author       = {Yaniv Leviathan and
                  Matan Kalman and
                  Yossi Matias},
  editor       = {Andreas Krause and
                  Emma Brunskill and
                  Kyunghyun Cho and
                  Barbara Engelhardt and
                  Sivan Sabato and
                  Jonathan Scarlett},
  title        = {Fast Inference from Transformers via Speculative Decoding},
  booktitle    = {International Conference on Machine Learning, {ICML} 2023, 23-29 July
                  2023, Honolulu, Hawaii, {USA}},
  series       = {Proceedings of Machine Learning Research},
  volume       = {202},
  pages        = {19274--19286},
  publisher    = {{PMLR}},
  year         = {2023},
  url          = {https://proceedings.mlr.press/v202/leviathan23a.html},
  bibsource    = {dblp computer science bibliography, https://dblp.org}
}

@inproceedings{loongserve,
  author    = {Bingyang Wu and
               Shengyu Liu and
               Yinmin Zhong and
               Peng Sun and
               Xuanzhe Liu and
               Xin Jin},
  editor    = {Emmett Witchel and
               Christopher J. Rossbach and
               Andrea C. Arpaci{-}Dusseau and
               Kimberly Keeton},
  title     = {LoongServe: Efficiently Serving Long-Context Large Language Models
               with Elastic Sequence Parallelism},
  booktitle = {Proceedings of the {ACM} {SIGOPS} 30th Symposium on Operating Systems
               Principles, {SOSP} 2024, Austin, TX, USA, November 4-6, 2024},
  pages     = {640--654},
  publisher = {{ACM}},
  year      = {2024},
  url       = {https://doi.org/10.1145/3694715.3695948},
  doi       = {10.1145/3694715.3695948},
  bibsource = {dblp computer science bibliography, https://dblp.org}
}

@inproceedings{DBLP:conf/nsdi/ZhongZWLCWHXMZ025,
  author       = {Yinmin Zhong and
                  Zili Zhang and
                  Bingyang Wu and
                  Shengyu Liu and
                  Yukun Chen and
                  Changyi Wan and
                  Hanpeng Hu and
                  Lei Xia and
                  Ranchen Ming and
                  Yibo Zhu and
                  Xin Jin},
  editor       = {Theophilus A. Benson and
                  Radhika Niranjan Mysore},
  title        = {Optimizing {RLHF} Training for Large Language Models with Stage Fusion},
  booktitle    = {22nd {USENIX} Symposium on Networked Systems Design and Implementation,
                  {NSDI} 2025, Philadelphia, PA, USA, April 28-30, 2025},
  pages        = {489--503},
  publisher    = {{USENIX} Association},
  year         = {2025},
  url          = {https://www.usenix.org/conference/nsdi25/presentation/zhong},
  bibsource    = {dblp computer science bibliography, https://dblp.org}
}

@article{realhf,
  author       = {Zhiyu Mei and
                  Wei Fu and
                  Kaiwei Li and
                  Guangju Wang and
                  Huanchen Zhang and
                  Yi Wu},
  title        = {ReaLHF: Optimized {RLHF} Training for Large Language Models through
                  Parameter Reallocation},
  journal      = {CoRR},
  volume       = {abs/2406.14088},
  year         = {2024},
  url          = {https://doi.org/10.48550/arXiv.2406.14088},
  doi          = {10.48550/ARXIV.2406.14088},
  eprinttype    = {arXiv},
  eprint       = {2406.14088},
  bibsource    = {dblp computer science bibliography, https://dblp.org}
}

@inproceedings{rlhfuse,
  author       = {Yinmin Zhong and
                  Zili Zhang and
                  Bingyang Wu and
                  Shengyu Liu and
                  Yukun Chen and
                  Changyi Wan and
                  Hanpeng Hu and
                  Lei Xia and
                  Ranchen Ming and
                  Yibo Zhu and
                  Xin Jin},
  editor       = {Theophilus A. Benson and
                  Radhika Niranjan Mysore},
  title        = {Optimizing {RLHF} Training for Large Language Models with Stage Fusion},
  booktitle    = {22nd {USENIX} Symposium on Networked Systems Design and Implementation,
                  {NSDI} 2025, Philadelphia, PA, USA, April 28-30, 2025},
  pages        = {489--503},
  publisher    = {{USENIX} Association},
  year         = {2025},
  url          = {https://www.usenix.org/conference/nsdi25/presentation/zhong},
  bibsource    = {dblp computer science bibliography, https://dblp.org}
}

@inproceedings{hybridflow,
  author       = {Guangming Sheng and
                  Chi Zhang and
                  Zilingfeng Ye and
                  Xibin Wu and
                  Wang Zhang and
                  Ru Zhang and
                  Yanghua Peng and
                  Haibin Lin and
                  Chuan Wu},
  title        = {HybridFlow: {A} Flexible and Efficient {RLHF} Framework},
  booktitle    = {Proceedings of the Twentieth European Conference on Computer Systems,
                  EuroSys 2025, Rotterdam, The Netherlands, 30 March 2025 - 3 April
                  2025},
  pages        = {1279--1297},
  publisher    = {{ACM}},
  year         = {2025},
  url          = {https://doi.org/10.1145/3689031.3696075},
  doi          = {10.1145/3689031.3696075},
  bibsource    = {dblp computer science bibliography, https://dblp.org}
}

@article{DBLP:journals/corr/abs-2501-12948,
  author       = {DeepSeek{-}AI and
                  Daya Guo and
                  Dejian Yang and
                  Haowei Zhang and
                  et. al},
  title        = {DeepSeek-R1: Incentivizing Reasoning Capability in LLMs via Reinforcement
                  Learning},
  journal      = {CoRR},
  volume       = {abs/2501.12948},
  year         = {2025},
  url          = {https://doi.org/10.48550/arXiv.2501.12948},
  doi          = {10.48550/ARXIV.2501.12948},
  eprinttype    = {arXiv},
  eprint       = {2501.12948},
  bibsource    = {dblp computer science bibliography, https://dblp.org}
}

@inproceedings{DBLP:conf/iclr/KeskarMNST17,
  author       = {Nitish Shirish Keskar and
                  Dheevatsa Mudigere and
                  Jorge Nocedal and
                  Mikhail Smelyanskiy and
                  Ping Tak Peter Tang},
  title        = {On Large-Batch Training for Deep Learning: Generalization Gap and
                  Sharp Minima},
  booktitle    = {5th International Conference on Learning Representations, {ICLR} 2017,
                  Toulon, France, April 24-26, 2017, Conference Track Proceedings},
  publisher    = {OpenReview.net},
  year         = {2017},
  url          = {https://openreview.net/forum?id=H1oyRlYgg},
  bibsource    = {dblp computer science bibliography, https://dblp.org}
}

@inproceedings{DBLP:conf/nips/YaoGLKM18,
  author       = {Zhewei Yao and
                  Amir Gholami and
                  Qi Lei and
                  Kurt Keutzer and
                  Michael W. Mahoney},
  editor       = {Samy Bengio and
                  Hanna M. Wallach and
                  Hugo Larochelle and
                  Kristen Grauman and
                  Nicol{\`{o}} Cesa{-}Bianchi and
                  Roman Garnett},
  title        = {Hessian-based Analysis of Large Batch Training and Robustness to Adversaries},
  booktitle    = {Advances in Neural Information Processing Systems 31: Annual Conference
                  on Neural Information Processing Systems 2018, NeurIPS 2018, December
                  3-8, 2018, Montr{\'{e}}al, Canada},
  pages        = {4954--4964},
  year         = {2018},
  url          = {https://proceedings.neurips.cc/paper/2018/hash/102f0bb6efb3a6128a3c750dd16729be-Abstract.html},
  bibsource    = {dblp computer science bibliography, https://dblp.org}
}

@article{DBLP:journals/corr/abs-2204-05862,
  author       = {Yuntao Bai and
                  Andy Jones and
                  Kamal Ndousse and
                  et.al},
  title        = {Training a Helpful and Harmless Assistant with Reinforcement Learning
                  from Human Feedback},
  journal      = {CoRR},
  volume       = {abs/2204.05862},
  year         = {2022},
  url          = {https://doi.org/10.48550/arXiv.2204.05862},
  doi          = {10.48550/ARXIV.2204.05862},
  eprinttype    = {arXiv},
  eprint       = {2204.05862},
  bibsource    = {dblp computer science bibliography, https://dblp.org}
}

@article{DBLP:journals/corr/abs-2303-08774,
  author       = {OpenAI},
  title        = {{GPT-4} Technical Report},
  journal      = {CoRR},
  volume       = {abs/2303.08774},
  year         = {2023},
  url          = {https://doi.org/10.48550/arXiv.2303.08774},
  doi          = {10.48550/ARXIV.2303.08774},
  eprinttype    = {arXiv},
  eprint       = {2303.08774},
  bibsource    = {dblp computer science bibliography, https://dblp.org}
}

@article{DBLP:journals/corr/abs-2112-09332,
  author       = {Reiichiro Nakano and
                  Jacob Hilton and
                  Suchir Balaji and
                  Jeff Wu and
                  Long Ouyang and
                  Christina Kim and
                  Christopher Hesse and
                  Shantanu Jain and
                  Vineet Kosaraju and
                  William Saunders and
                  Xu Jiang and
                  Karl Cobbe and
                  Tyna Eloundou and
                  Gretchen Krueger and
                  Kevin Button and
                  Matthew Knight and
                  Benjamin Chess and
                  John Schulman},
  title        = {WebGPT: Browser-assisted question-answering with human feedback},
  journal      = {CoRR},
  volume       = {abs/2112.09332},
  year         = {2021},
  url          = {https://arxiv.org/abs/2112.09332},
  eprinttype    = {arXiv},
  eprint       = {2112.09332},
  bibsource    = {dblp computer science bibliography, https://dblp.org}
}

@article{DBLP:journals/corr/abs-2308-12950,
  author       = {Baptiste Rozi{\`{e}}re and
                  Jonas Gehring and
                  Fabian Gloeckle and
                  Sten Sootla and
                  Itai Gat and
                  Xiaoqing Ellen Tan and
                  Yossi Adi and
                  Jingyu Liu and
                  Tal Remez and
                  J{\'{e}}r{\'{e}}my Rapin and
                  Artyom Kozhevnikov and
                  Ivan Evtimov and
                  Joanna Bitton and
                  Manish Bhatt and
                  Cristian Canton{-}Ferrer and
                  Aaron Grattafiori and
                  Wenhan Xiong and
                  Alexandre D{\'{e}}fossez and
                  Jade Copet and
                  Faisal Azhar and
                  Hugo Touvron and
                  Louis Martin and
                  Nicolas Usunier and
                  Thomas Scialom and
                  Gabriel Synnaeve},
  title        = {Code Llama: Open Foundation Models for Code},
  journal      = {CoRR},
  volume       = {abs/2308.12950},
  year         = {2023},
  url          = {https://doi.org/10.48550/arXiv.2308.12950},
  doi          = {10.48550/ARXIV.2308.12950},
  eprinttype    = {arXiv},
  eprint       = {2308.12950},
  bibsource    = {dblp computer science bibliography, https://dblp.org}
}

@article{DBLP:journals/corr/abs-2402-03300,
  author       = {Zhihong Shao and
                  Peiyi Wang and
                  Qihao Zhu and
                  Runxin Xu and
                  Junxiao Song and
                  Mingchuan Zhang and
                  Y. K. Li and
                  Y. Wu and
                  Daya Guo},
  title        = {DeepSeekMath: Pushing the Limits of Mathematical Reasoning in Open
                  Language Models},
  journal      = {CoRR},
  volume       = {abs/2402.03300},
  year         = {2024},
  url          = {https://doi.org/10.48550/arXiv.2402.03300},
  doi          = {10.48550/ARXIV.2402.03300},
  eprinttype    = {arXiv},
  eprint       = {2402.03300},
  bibsource    = {dblp computer science bibliography, https://dblp.org}
}

@article{turbospec,
      author={Xiaoxuan Liu and Jongseok Park and Langxiang Hu and Woosuk Kwon and Zhuohan Li and Chen Zhang and Kuntai Du and Xiangxi Mo and Kaichao You and Alvin Cheung and Zhijie Deng and Ion Stoica and Hao Zhang},
      title={TurboSpec: Closed-loop Speculation Control System for Optimizing LLM Serving Goodput}, 
      year={2025},
      journal      = {CoRR},
      eprinttype    = {arXiv},
      eprint={2406.14066},
      archivePrefix={arXiv},
      primaryClass={cs.AI},
      url={https://arxiv.org/abs/2406.14066}, 
}

@misc{gspo,
      author={Chujie Zheng and Shixuan Liu and Mingze Li and Xiong-Hui Chen and Bowen Yu and Chang Gao and Kai Dang and Yuqiong Liu and Rui Men and An Yang and Jingren Zhou and Junyang Lin},
      title={Group Sequence Policy Optimization}, 
      year={2025},
      journal      = {CoRR},
      eprinttype    = {arXiv},
      eprint={2507.18071},
      archivePrefix={arXiv},
      primaryClass={cs.LG},
      url={https://arxiv.org/abs/2507.18071}, 
}

@article{DBLP:journals/corr/abs-2302-01318,
  author       = {Charlie Chen and
                  Sebastian Borgeaud and
                  Geoffrey Irving and
                  Jean{-}Baptiste Lespiau and
                  Laurent Sifre and
                  John Jumper},
  title        = {Accelerating Large Language Model Decoding with Speculative Sampling},
  journal      = {CoRR},
  volume       = {abs/2302.01318},
  year         = {2023},
  url          = {https://doi.org/10.48550/arXiv.2302.01318},
  doi          = {10.48550/ARXIV.2302.01318},
  eprinttype    = {arXiv},
  eprint       = {2302.01318},
  bibsource    = {dblp computer science bibliography, https://dblp.org}
}

@article{swiftspec,
  author       = {Ziyi Zhang and
                  Ziheng Jiang and
                  Chengquan Jiang and
                  Menghan Yu and
                  Size Zheng and
                  Haibin Lin and
                  Henry Hoffmann and
                  Xin Liu},
  title        = {SwiftSpec: Ultra-Low Latency {LLM} Decoding by Scaling Asynchronous
                  Speculative Decoding},
  journal      = {CoRR},
  volume       = {abs/2506.11309},
  year         = {2025},
  url          = {https://doi.org/10.48550/arXiv.2506.11309},
  doi          = {10.48550/ARXIV.2506.11309},
  eprinttype    = {arXiv},
  eprint       = {2506.11309},
  bibsource    = {dblp computer science bibliography, https://dblp.org}
}

@inproceedings{pearl,
  author       = {Tianyu Liu and
                  Yun Li and
                  Qitan Lv and
                  Kai Liu and
                  Jianchen Zhu and
                  Winston Hu and
                  Xiao Sun},
  title        = {{PEARL:} Parallel Speculative Decoding with Adaptive Draft Length},
  booktitle    = {The Thirteenth International Conference on Learning Representations,
                  {ICLR} 2025, Singapore, April 24-28, 2025},
  publisher    = {OpenReview.net},
  year         = {2025},
  url          = {https://openreview.net/forum?id=QOXrVMiHGK},
  bibsource    = {dblp computer science bibliography, https://dblp.org}
}

@article{eagle-3,
  author       = {Yuhui Li and
                  Fangyun Wei and
                  Chao Zhang and
                  Hongyang Zhang},
  title        = {{EAGLE-3:} Scaling up Inference Acceleration of Large Language Models
                  via Training-Time Test},
  journal      = {CoRR},
  volume       = {abs/2503.01840},
  year         = {2025},
  url          = {https://doi.org/10.48550/arXiv.2503.01840},
  doi          = {10.48550/ARXIV.2503.01840},
  eprinttype    = {arXiv},
  eprint       = {2503.01840},
  bibsource    = {dblp computer science bibliography, https://dblp.org}
}

@inproceedings{eagle-2,
  author       = {Yuhui Li and
                  Fangyun Wei and
                  Chao Zhang and
                  Hongyang Zhang},
  editor       = {Yaser Al{-}Onaizan and
                  Mohit Bansal and
                  Yun{-}Nung Chen},
  title        = {{EAGLE-2:} Faster Inference of Language Models with Dynamic Draft
                  Trees},
  booktitle    = {Proceedings of the 2024 Conference on Empirical Methods in Natural
                  Language Processing, {EMNLP} 2024, Miami, FL, USA, November 12-16,
                  2024},
  pages        = {7421--7432},
  publisher    = {Association for Computational Linguistics},
  year         = {2024},
  url          = {https://doi.org/10.18653/v1/2024.emnlp-main.422},
  doi          = {10.18653/V1/2024.EMNLP-MAIN.422},
  bibsource    = {dblp computer science bibliography, https://dblp.org}
}

@article{dapo,
  author       = {Qiying Yu and
                  Zheng Zhang and
                  Ruofei Zhu and
                  Yufeng Yuan and
                  Xiaochen Zuo and
                  Yu Yue and
                  Tiantian Fan and
                  Gaohong Liu and
                  Lingjun Liu and
                  Xin Liu and
                  Haibin Lin and
                  Zhiqi Lin and
                  Bole Ma and
                  Guangming Sheng and
                  Yuxuan Tong and
                  Chi Zhang and
                  Mofan Zhang and
                  Wang Zhang and
                  Hang Zhu and
                  Jinhua Zhu and
                  Jiaze Chen and
                  Jiangjie Chen and
                  Chengyi Wang and
                  Hongli Yu and
                  Weinan Dai and
                  Yuxuan Song and
                  Xiangpeng Wei and
                  Hao Zhou and
                  Jingjing Liu and
                  Wei{-}Ying Ma and
                  Ya{-}Qin Zhang and
                  Lin Yan and
                  Mu Qiao and
                  Yonghui Wu and
                  Mingxuan Wang},
  title        = {{DAPO:} An Open-Source {LLM} Reinforcement Learning System at Scale},
  journal      = {CoRR},
  volume       = {abs/2503.14476},
  year         = {2025},
  url          = {https://doi.org/10.48550/arXiv.2503.14476},
  doi          = {10.48550/ARXIV.2503.14476},
  eprinttype    = {arXiv},
  eprint       = {2503.14476},
  bibsource    = {dblp computer science bibliography, https://dblp.org}
}

@article{DBLP:journals/corr/abs-2504-05118,
  author       = {Yu Yue and
                  Yufeng Yuan and
                  Qiying Yu and
                  Xiaochen Zuo and
                  Ruofei Zhu and
                  Wenyuan Xu and
                  Jiaze Chen and
                  Cheng{-}Xiang Wang and
                  Tiantian Fan and
                  Zhengyin Du and
                  Xiangpeng Wei and
                  Xiangyu Yu and
                  Gaohong Liu and
                  Juncai Liu and
                  Lingjun Liu and
                  Haibin Lin and
                  Zhiqi Lin and
                  Bole Ma and
                  Chi Zhang and
                  Mofan Zhang and
                  Wang Zhang and
                  Hang Zhu and
                  Ru Zhang and
                  Xin Liu and
                  Mingxuan Wang and
                  Yonghui Wu and
                  Lin Yan},
  title        = {{VAPO:} Efficient and Reliable Reinforcement Learning for Advanced
                  Reasoning Tasks},
  journal      = {CoRR},
  volume       = {abs/2504.05118},
  year         = {2025},
  url          = {https://doi.org/10.48550/arXiv.2504.05118},
  doi          = {10.48550/ARXIV.2504.05118},
  eprinttype    = {arXiv},
  eprint       = {2504.05118},
  bibsource    = {dblp computer science bibliography, https://dblp.org}
}

@inproceedings{DBLP:conf/nips/ShahBZTRD24,
  author       = {Jay Shah and
                  Ganesh Bikshandi and
                  Ying Zhang and
                  Vijay Thakkar and
                  Pradeep Ramani and
                  Tri Dao},
  editor       = {Amir Globersons and
                  Lester Mackey and
                  Danielle Belgrave and
                  Angela Fan and
                  Ulrich Paquet and
                  Jakub M. Tomczak and
                  Cheng Zhang},
  title        = {FlashAttention-3: Fast and Accurate Attention with Asynchrony and
                  Low-precision},
  booktitle    = {Advances in Neural Information Processing Systems 38: Annual Conference
                  on Neural Information Processing Systems 2024, NeurIPS 2024, Vancouver,
                  BC, Canada, December 10 - 15, 2024},
  year         = {2024},
  url          = {http://papers.nips.cc/paper\_files/paper/2024/hash/7ede97c3e082c6df10a8d6103a2eebd2-Abstract-Conference.html},
  bibsource    = {dblp computer science bibliography, https://dblp.org}
}

@inproceedings{DBLP:conf/eurosys/ShengZYWZZPL025,
  author       = {Guangming Sheng and
                  Chi Zhang and
                  Zilingfeng Ye and
                  Xibin Wu and
                  Wang Zhang and
                  Ru Zhang and
                  Yanghua Peng and
                  Haibin Lin and
                  Chuan Wu},
  title        = {HybridFlow: {A} Flexible and Efficient {RLHF} Framework},
  booktitle    = {Proceedings of the Twentieth European Conference on Computer Systems,
                  EuroSys 2025, Rotterdam, The Netherlands, 30 March 2025 - 3 April
                  2025},
  pages        = {1279--1297},
  publisher    = {{ACM}},
  year         = {2025},
  url          = {https://doi.org/10.1145/3689031.3696075},
  doi          = {10.1145/3689031.3696075},
  bibsource    = {dblp computer science bibliography, https://dblp.org}
}

@inproceedings{specinfer,
  author       = {Xupeng Miao and
                  Gabriele Oliaro and
                  Zhihao Zhang and
                  Xinhao Cheng and
                  Zeyu Wang and
                  Zhengxin Zhang and
                  Rae Ying Yee Wong and
                  Alan Zhu and
                  Lijie Yang and
                  Xiaoxiang Shi and
                  Chunan Shi and
                  Zhuoming Chen and
                  Daiyaan Arfeen and
                  Reyna Abhyankar and
                  Zhihao Jia},
  editor       = {Rajiv Gupta and
                  Nael B. Abu{-}Ghazaleh and
                  Madan Musuvathi and
                  Dan Tsafrir},
  title        = {SpecInfer: Accelerating Large Language Model Serving with Tree-based
                  Speculative Inference and Verification},
  booktitle    = {Proceedings of the 29th {ACM} International Conference on Architectural
                  Support for Programming Languages and Operating Systems, Volume 3,
                  {ASPLOS} 2024, La Jolla, CA, USA, 27 April 2024- 1 May 2024},
  pages        = {932--949},
  publisher    = {{ACM}},
  year         = {2024},
  url          = {https://doi.org/10.1145/3620666.3651335},
  doi          = {10.1145/3620666.3651335},
  bibsource    = {dblp computer science bibliography, https://dblp.org}
}

@inproceedings{spd,
  author       = {Yaniv Leviathan and
                  Matan Kalman and
                  Yossi Matias},
  editor       = {Andreas Krause and
                  Emma Brunskill and
                  Kyunghyun Cho and
                  Barbara Engelhardt and
                  Sivan Sabato and
                  Jonathan Scarlett},
  title        = {Fast Inference from Transformers via Speculative Decoding},
  booktitle    = {International Conference on Machine Learning, {ICML} 2023, 23-29 July
                  2023, Honolulu, Hawaii, {USA}},
  series       = {Proceedings of Machine Learning Research},
  volume       = {202},
  pages        = {19274--19286},
  publisher    = {{PMLR}},
  year         = {2023},
  url          = {https://proceedings.mlr.press/v202/leviathan23a.html},
  bibsource    = {dblp computer science bibliography, https://dblp.org}
}

@article{streamrl,
  author       = {Yinmin Zhong and
                  Zili Zhang and
                  Xiaoniu Song and
                  Hanpeng Hu and
                  Chao Jin and
                  Bingyang Wu and
                  Nuo Chen and
                  Yukun Chen and
                  Yu Zhou and
                  Changyi Wan and
                  Hongyu Zhou and
                  Yimin Jiang and
                  Yibo Zhu and
                  Daxin Jiang},
  title        = {StreamRL: Scalable, Heterogeneous, and Elastic {RL} for LLMs with
                  Disaggregated Stream Generation},
  journal      = {CoRR},
  volume       = {abs/2504.15930},
  year         = {2025},
  url          = {https://doi.org/10.48550/arXiv.2504.15930},
  doi          = {10.48550/ARXIV.2504.15930},
  eprinttype    = {arXiv},
  eprint       = {2504.15930},
  bibsource    = {dblp computer science bibliography, https://dblp.org}
}

@article{ulysses-sp,
  author       = {Sam Ade Jacobs and
                  Masahiro Tanaka and
                  Chengming Zhang and
                  Minjia Zhang and
                  Shuaiwen Leon Song and
                  Samyam Rajbhandari and
                  Yuxiong He},
  title        = {DeepSpeed Ulysses: System Optimizations for Enabling Training of Extreme
                  Long Sequence Transformer Models},
  journal      = {CoRR},
  volume       = {abs/2309.14509},
  year         = {2023},
  url          = {https://doi.org/10.48550/arXiv.2309.14509},
  doi          = {10.48550/ARXIV.2309.14509},
  eprinttype    = {arXiv},
  eprint       = {2309.14509},
  bibsource    = {dblp computer science bibliography, https://dblp.org}
}

@article{DBLP:journals/corr/abs-2505-19645,
  author       = {Zongle Huang and
                  Lei Zhu and
                  Zongyuan Zhan and
                  Ting Hu and
                  Weikai Mao and
                  Xianzhi Yu and
                  Yongpan Liu and
                  Tianyu Zhang},
  title        = {MoESD: Unveil Speculative Decoding's Potential for Accelerating
                  Sparse MoE},
  journal      = {CoRR},
  volume       = {abs/2505.19645},
  year         = {2025},
  url          = {https://doi.org/10.48550/arXiv.2505.19645},
  doi          = {10.48550/ARXIV.2505.19645},
  eprinttype    = {arXiv},
  eprint       = {2505.19645},
  bibsource    = {dblp computer science bibliography, https://dblp.org}
}

@article{kunserve,
  author       = {Rongxin Cheng and
                  Yuxin Lai and
                  Xingda Wei and
                  Rong Chen and
                  Haibo Chen},
  title        = {KunServe: Parameter-centric Memory Management for Efficient Memory Overloading Handling in LLM Serving},
  journal      = {CoRR},
  volume       = {abs/2412.18169},
  year         = {2024},
  url          = {https://doi.org/10.48550/arXiv.2412.18169},
  doi          = {10.48550/ARXIV.2412.18169},
  eprinttype    = {arXiv},
  eprint       = {2412.18169},
  bibsource    = {dblp computer science bibliography, https://dblp.org}
}

@misc{liu2025specrlacceleratingonpolicyreinforcement,
      title={SPEC-RL: Accelerating On-Policy Reinforcement Learning via Speculative Rollouts}, 
      author={Bingshuai Liu and Ante Wang and Zijun Min and Liang Yao and Haibo Zhang and Yang Liu and Anxiang Zeng and Jinsong Su},
      year={2025},
      eprint={2509.23232},
      archivePrefix={arXiv},
      primaryClass={cs.LG},
      url={https://arxiv.org/abs/2509.23232}, 
}

@article{deepseek-r1,
  author     = {Daya Guo, Dejian Yang, Haowei Zhang et al.},
  title      = {DeepSeek-R1 incentivizes reasoning in LLMs through reinforcement learning},
  journal    = {Nature},
  volume     = {645},
  pages      = {633--638},
  year       = {2025},
  url        = {https://doi.org/10.1038/s41586-025-09422-z},
  doi        = {10.1038/s41586-025-09422-z},
}

@article{DBLP:journals/corr/abs-2504-02605,
  author       = {Daoguang Zan and
                  Zhirong Huang and
                  Wei Liu and
                  Hanwu Chen and
                  Linhao Zhang and
                  Shulin Xin and
                  Lu Chen and
                  Qi Liu and
                  Xiaojian Zhong and
                  Aoyan Li and
                  Siyao Liu and
                  Yongsheng Xiao and
                  Liangqiang Chen and
                  Yuyu Zhang and
                  Jing Su and
                  Tianyu Liu and
                  Rui Long and
                  Kai Shen and
                  Liang Xiang},
  title        = {Multi-SWE-bench: {A} Multilingual Benchmark for Issue Resolving},
  journal      = {CoRR},
  volume       = {abs/2504.02605},
  year         = {2025},
  url          = {https://doi.org/10.48550/arXiv.2504.02605},
  doi          = {10.48550/ARXIV.2504.02605},
  eprinttype    = {arXiv},
  eprint       = {2504.02605},
  bibsource    = {dblp computer science bibliography, https://dblp.org}
}

@article{DBLP:journals/corr/abs-2509-02547,
  author       = {Guibin Zhang and
                  Hejia Geng and
                  Xiaohang Yu and
                  Zhenfei Yin and
                  Zaibin Zhang and
                  Zelin Tan and
                  Heng Zhou and
                  Zhongzhi Li and
                  Xiangyuan Xue and
                  Yijiang Li and
                  Yifan Zhou and
                  Yang Chen and
                  Chen Zhang and
                  Yutao Fan and
                  Zihu Wang and
                  Songtao Huang and
                  Yue Liao and
                  Hongru Wang and
                  Mengyue Yang and
                  Heng Ji and
                  Michael Littman and
                  Jun Wang and
                  Shuicheng Yan and
                  Philip Torr and
                  Lei Bai},
  title        = {The Landscape of Agentic Reinforcement Learning for LLMs: {A} Survey},
  journal      = {CoRR},
  volume       = {abs/2509.02547},
  year         = {2025},
  url          = {https://doi.org/10.48550/arXiv.2509.02547},
  doi          = {10.48550/ARXIV.2509.02547},
  eprinttype    = {arXiv},
  eprint       = {2509.02547},
  bibsource    = {dblp computer science bibliography, https://dblp.org}
}

@article{DBLP:journals/corr/abs-2506-20520,
  author       = {Charles Arnal and
                  Ga{\"{e}}tan Narozniak and
                  Vivien Cabannes and
                  Yunhao Tang and
                  Julia Kempe and
                  R{\'{e}}mi Munos},
  title        = {Asymmetric {REINFORCE} for off-Policy Reinforcement Learning: Balancing
                  positive and negative rewards},
  journal      = {CoRR},
  volume       = {abs/2506.20520},
  year         = {2025},
  url          = {https://doi.org/10.48550/arXiv.2506.20520},
  doi          = {10.48550/ARXIV.2506.20520},
  eprinttype    = {arXiv},
  eprint       = {2506.20520},
  bibsource    = {dblp computer science bibliography, https://dblp.org}
}

@article{DBLP:journals/corr/abs-2503-14286,
  author       = {Nicolas Le Roux and
                  Marc G. Bellemare and
                  Jonathan Lebensold and
                  Arnaud Bergeron and
                  Joshua Greaves and
                  Alexandre Fr{\'{e}}chette and
                  Carolyne Pelletier and
                  Eric Thibodeau{-}Laufer and
                  S{\'{a}}ndor T{\'{o}}th and
                  Sam Work},
  title        = {Tapered Off-Policy {REINFORCE:} Stable and efficient reinforcement
                  learning for LLMs},
  journal      = {CoRR},
  volume       = {abs/2503.14286},
  year         = {2025},
  url          = {https://doi.org/10.48550/arXiv.2503.14286},
  doi          = {10.48550/ARXIV.2503.14286},
  eprinttype    = {arXiv},
  eprint       = {2503.14286},
  bibsource    = {dblp computer science bibliography, https://dblp.org}
}

@article{DBLP:journals/corr/abs-2405-08448,
  author       = {Yunhao Tang and
                  Zhaohan Daniel Guo and
                  Zeyu Zheng and
                  Daniele Calandriello and
                  Yuan Cao and
                  Eugene Tarassov and
                  R{\'{e}}mi Munos and
                  Bernardo {\'{A}}vila Pires and
                  Michal Valko and
                  Yong Cheng and
                  Will Dabney},
  title        = {Understanding the performance gap between online and offline alignment
                  algorithms},
  journal      = {CoRR},
  volume       = {abs/2405.08448},
  year         = {2024},
  url          = {https://doi.org/10.48550/arXiv.2405.08448},
  doi          = {10.48550/ARXIV.2405.08448},
  eprinttype    = {arXiv},
  eprint       = {2405.08448},
  bibsource    = {dblp computer science bibliography, https://dblp.org}
}

@article{rhymerl,
  author       = {Jingkai He and
                  Tianjian Li and
                  Erhu Feng and
                  Dong Du and
                  Qian Liu and
                  Tao Liu and
                  Yubin Xia and
                  Haibo Chen},
  title        = {History Rhymes: Accelerating {LLM} Reinforcement Learning with RhymeRL},
  journal      = {CoRR},
  volume       = {abs/2508.18588},
  year         = {2025},
  url          = {https://doi.org/10.48550/arXiv.2508.18588},
  doi          = {10.48550/ARXIV.2508.18588},
  eprinttype    = {arXiv},
  eprint       = {2508.18588},
  bibsource    = {dblp computer science bibliography, https://dblp.org}
}

@article{rollpacker,
  author       = {Wei Gao and
                  Yuheng Zhao and
                  Dakai An and
                  Tianyuan Wu and
                  Lunxi Cao and
                  Shaopan Xiong and
                  Ju Huang and
                  Weixun Wang and
                  Siran Yang and
                  Wenbo Su and
                  Jiamang Wang and
                  Lin Qu and
                  Bo Zheng and
                  Wei Wang},
  title        = {RollPacker: Mitigating Long-Tail Rollouts for Fast, Synchronous {RL}
                  Post-Training},
  journal      = {CoRR},
  volume       = {abs/2509.21009},
  year         = {2025},
  url          = {https://doi.org/10.48550/arXiv.2509.21009},
  doi          = {10.48550/ARXIV.2509.21009},
  eprinttype    = {arXiv},
  eprint       = {2509.21009},
  bibsource    = {dblp computer science bibliography, https://dblp.org}
}

@inproceedings{DBLP:conf/icml/LiuHBCDS024,
  author       = {Xiaoxuan Liu and
                  Lanxiang Hu and
                  Peter Bailis and
                  Alvin Cheung and
                  Zhijie Deng and
                  Ion Stoica and
                  Hao Zhang},
  title        = {Online Speculative Decoding},
  booktitle    = {Forty-first International Conference on Machine Learning, {ICML} 2024,
                  Vienna, Austria, July 21-27, 2024},
  publisher    = {OpenReview.net},
  year         = {2024},
  url          = {https://openreview.net/forum?id=BPQHXwVNvl},
  bibsource    = {dblp computer science bibliography, https://dblp.org}
}

@article{rlboost,
  author       = {Eloy Anguiano Batanero and
                  {\'{A}}ngela Fern{\'{a}}ndez Pascual and
                  {\'{A}}lvaro Barbero Jim{\'{e}}nez},
  title        = {RLBoost: Boosting supervised models using deep reinforcement learning},
  journal      = {Neurocomputing},
  volume       = {618},
  pages        = {128815},
  year         = {2025},
  url          = {https://doi.org/10.1016/j.neucom.2024.128815},
  doi          = {10.1016/J.NEUCOM.2024.128815},
  bibsource    = {dblp computer science bibliography, https://dblp.org}
}

@article{DBLP:journals/corr/abs-2501-12599,
  author       = {Kimi Team and
                  Angang Du and
                  Bofei Gao and et.al},
  title        = {Kimi k1.5: Scaling Reinforcement Learning with LLMs},
  journal      = {CoRR},
  volume       = {abs/2501.12599},
  year         = {2025},
  url          = {https://doi.org/10.48550/arXiv.2501.12599},
  doi          = {10.48550/ARXIV.2501.12599},
  eprinttype    = {arXiv},
  eprint       = {2501.12599},
  bibsource    = {dblp computer science bibliography, https://dblp.org}
}

@article{DBLP:journals/corr/abs-2507-20534,
  author       = {Yifan Bai and
                  Yiping Bao and
                  Guanduo Chen and et. al
                  },
  title        = {Kimi {K2:} Open Agentic Intelligence},
  journal      = {CoRR},
  volume       = {abs/2507.20534},
  year         = {2025},
  url          = {https://doi.org/10.48550/arXiv.2507.20534},
  doi          = {10.48550/ARXIV.2507.20534},
  eprinttype    = {arXiv},
  eprint       = {2507.20534},
  bibsource    = {dblp computer science bibliography, https://dblp.org}
}

@article{DBLP:journals/corr/abs-2505-23433,
  author       = {Jian Yao and
                  Ran Cheng and
                  Xingyu Wu and
                  Jibin Wu and
                  Kay Chen Tan},
  title        = {Diversity-Aware Policy Optimization for Large Language Model Reasoning},
  journal      = {CoRR},
  volume       = {abs/2505.23433},
  year         = {2025},
  url          = {https://doi.org/10.48550/arXiv.2505.23433},
  doi          = {10.48550/ARXIV.2505.23433},
  eprinttype    = {arXiv},
  eprint       = {2505.23433},
  bibsource    = {dblp computer science bibliography, https://dblp.org}
}

@article{DBLP:journals/corr/abs-2508-03501,
  author       = {Alexander Golubev and
                  Maria Trofimova and
                  Sergei Polezhaev and
                  Ibragim Badertdinov and
                  Maksim Nekrashevich and
                  Anton Shevtsov and
                  Simon Karasik and
                  Sergey Abramov and
                  Andrei Andriushchenko and
                  Filipp Fisin and
                  Sergei Skvortsov and
                  Boris Yangel},
  title        = {Training Long-Context, Multi-Turn Software Engineering Agents with
                  Reinforcement Learning},
  journal      = {CoRR},
  volume       = {abs/2508.03501},
  year         = {2025},
  url          = {https://doi.org/10.48550/arXiv.2508.03501},
  doi          = {10.48550/ARXIV.2508.03501},
  eprinttype    = {arXiv},
  eprint       = {2508.03501},
  bibsource    = {dblp computer science bibliography, https://dblp.org}
}

@article{DBLP:journals/corr/abs-2509-04474,
  author       = {Shengyin Sun and
                  Yiming Li and
                  Xing Li and
                  Yingzhao Lian and
                  Weizhe Lin and
                  Hui{-}Ling Zhen and
                  Zhiyuan Yang and
                  Chen Chen and
                  Xianzhi Yu and
                  Mingxuan Yuan and
                  Chen Ma},
  title        = {Scaling Up, Speeding Up: {A} Benchmark of Speculative Decoding for
                  Efficient {LLM} Test-Time Scaling},
  journal      = {CoRR},
  volume       = {abs/2509.04474},
  year         = {2025},
  url          = {https://doi.org/10.48550/arXiv.2509.04474},
  doi          = {10.48550/ARXIV.2509.04474},
  eprinttype    = {arXiv},
  eprint       = {2509.04474},
  bibsource    = {dblp computer science bibliography, https://dblp.org}
}

@misc{pld,
  title       = {Prompt lookup decoding},
  author      = {Apoorv Saxena, Arpit Tarang Saxena},
  howpublished = {\burl{https://github.com/apoorvumang/prompt-lookup-decoding}},
  year = {2025},
}

@inproceedings{lookahead,
  author       = {Yichao Fu and
                  Peter Bailis and
                  Ion Stoica and
                  Hao Zhang},
  title        = {Break the Sequential Dependency of {LLM} Inference Using Lookahead
                  Decoding},
  booktitle    = {Forty-first International Conference on Machine Learning, {ICML} 2024,
                  Vienna, Austria, July 21-27, 2024},
  publisher    = {OpenReview.net},
  year         = {2024},
  url          = {https://openreview.net/forum?id=eDjvSFOkXw},
  bibsource    = {dblp computer science bibliography, https://dblp.org}
}

@article{DBLP:journals/corr/abs-2412-19437,
  author       = {DeepSeek{-}AI},
  title        = {DeepSeek-V3 Technical Report},
  journal      = {CoRR},
  volume       = {abs/2412.19437},
  year         = {2024},
  url          = {https://doi.org/10.48550/arXiv.2412.19437},
  doi          = {10.48550/ARXIV.2412.19437},
  eprinttype    = {arXiv},
  eprint       = {2412.19437},
  bibsource    = {dblp computer science bibliography, https://dblp.org}
}

@inproceedings{swebench,
  author       = {Carlos E. Jimenez and
                  John Yang and
                  Alexander Wettig and
                  Shunyu Yao and
                  Kexin Pei and
                  Ofir Press and
                  Karthik R. Narasimhan},
  title        = {SWE-bench: Can Language Models Resolve Real-world Github Issues?},
  booktitle    = {The Twelfth International Conference on Learning Representations,
                  {ICLR} 2024, Vienna, Austria, May 7-11, 2024},
  publisher    = {OpenReview.net},
  year         = {2024},
  url          = {https://openreview.net/forum?id=VTF8yNQM66},
  bibsource    = {dblp computer science bibliography, https://dblp.org}
}

@inproceedings{agentbench,
  author       = {Xiao Liu and
                  Hao Yu and
                  Hanchen Zhang and
                  et. al},
  title        = {AgentBench: Evaluating LLMs as Agents},
  booktitle    = {The Twelfth International Conference on Learning Representations,
                  {ICLR} 2024, Vienna, Austria, May 7-11, 2024},
  publisher    = {OpenReview.net},
  year         = {2024},
  url          = {https://openreview.net/forum?id=zAdUB0aCTQ},
  bibsource    = {dblp computer science bibliography, https://dblp.org}
}

@article{DBLP:journals/corr/abs-2505-23585,
  author       = {Yaru Hao and
                  Li Dong and
                  Xun Wu and
                  Shaohan Huang and
                  Zewen Chi and
                  Furu Wei},
  title        = {On-Policy {RL} with Optimal Reward Baseline},
  journal      = {CoRR},
  volume       = {abs/2505.23585},
  year         = {2025},
  url          = {https://doi.org/10.48550/arXiv.2505.23585},
  doi          = {10.48550/ARXIV.2505.23585},
  eprinttype    = {arXiv},
  eprint       = {2505.23585},
  bibsource    = {dblp computer science bibliography, https://dblp.org}
}

@article{seer,
  author       = {Ruoyu Qin and
                  Weiran He and
                  Weixiao Huang and
                  Yangkun Zhang and
                  Yikai Zhao and
                  Bo Pang and
                  Xinran Xu and
                  Yingdi Shan and
                  Yongwei Wu and
                  Mingxing Zhang},
  title        = {Seer: Online Context Learning for Fast Synchronous LLM Reinforcement Learning},
  journal      = {CoRR},
  volume       = {abs/2511.14617},
  year         = {2025},
  url          = {https://arxiv.org/pdf/2511.14617},
  doi          = {10.48550/ARXIV.2511.14617},
  eprinttype    = {arXiv},
  eprint       = {2511.14617},
}

@inproceedings{eagle,
  author       = {Yuhui Li and
                  Fangyun Wei and
                  Chao Zhang and
                  Hongyang Zhang},
  title        = {{EAGLE:} Speculative Sampling Requires Rethinking Feature Uncertainty},
  booktitle    = {Forty-first International Conference on Machine Learning, {ICML} 2024,
                  Vienna, Austria, July 21-27, 2024},
  publisher    = {OpenReview.net},
  year         = {2024},
  url          = {https://openreview.net/forum?id=1NdN7eXyb4},
  bibsource    = {dblp computer science bibliography, https://dblp.org}
}

@inproceedings{DBLP:conf/acl/HuWZZLCZ25,
  author       = {Yuxuan Hu and
                  Ke Wang and
                  Xiaokang Zhang and
                  Fanjin Zhang and
                  Cuiping Li and
                  Hong Chen and
                  Jing Zhang},
  editor       = {Wanxiang Che and
                  Joyce Nabende and
                  Ekaterina Shutova and
                  Mohammad Taher Pilehvar},
  title        = {{SAM} Decoding: Speculative Decoding via Suffix Automaton},
  booktitle    = {Proceedings of the 63rd Annual Meeting of the Association for Computational
                  Linguistics (Volume 1: Long Papers), {ACL} 2025, Vienna, Austria,
                  July 27 - August 1, 2025},
  pages        = {12187--12204},
  publisher    = {Association for Computational Linguistics},
  year         = {2025},
  url          = {https://aclanthology.org/2025.acl-long.595/},
  bibsource    = {dblp computer science bibliography, https://dblp.org}
}

@article{tlt,
  author       = {Qinghao Hu and
                  Shang Yang and
                  Junxian Guo and
                  Xiaozhe Yao and
                  Yujun Lin and
                  Yuxian Gu and
                  Han Cai and
                  Chuang Gan and
                  Ana Klimovic and
                  Song Han},
  title        = {Taming the Long-Tail: Efficient Reasoning RL Training with Adaptive Drafter},
  journal      = {CoRR},
  volume       = {abs/2511.16665},
  year         = {2025},
  url          = {https://arxiv.org/pdf/2511.16665},
  doi          = {10.48550/ARXIV.2511.16665},
  eprinttype    = {arXiv},
  eprint       = {2511.16665},
}

@article{DBLP:journals/corr/abs-2504-13914,
  author       = {Jiaze Chen and
                  Tiantian Fan and
                  Xin Liu and
                  et. al},
  title        = {Seed1.5-Thinking: Advancing Superb Reasoning Models with Reinforcement
                  Learning},
  journal      = {CoRR},
  volume       = {abs/2504.13914},
  year         = {2025},
  url          = {https://doi.org/10.48550/arXiv.2504.13914},
  doi          = {10.48550/ARXIV.2504.13914},
  eprinttype    = {arXiv},
  eprint       = {2504.13914},
  bibsource    = {dblp computer science bibliography, https://dblp.org}
}

@article{qwen3,
  author       = {An Yang and
                  Anfeng Li and
                  Baosong Yang and et. al},
  title        = {Qwen3 Technical Report},
  journal      = {CoRR},
  volume       = {abs/2505.09388},
  year         = {2025},
  url          = {https://doi.org/10.48550/arXiv.2505.09388},
  doi          = {10.48550/ARXIV.2505.09388},
  eprinttype    = {arXiv},
  eprint       = {2505.09388},
  bibsource    = {dblp computer science bibliography, https://dblp.org}
}

@inproceedings{perf-agent,
  author       = {Dan Williams and
                  Milo Craun and
                  Michael V. Le and
                  Julian James Stephen and
                  Salman Ahmed and
                  Hani Jamjoom},
  title        = {Towards Safe Agentic {AI} Performance Engineering},
  booktitle    = {Proceedings of the 4th Workshop on Practical Adoption Challenges of
                  {ML} for Systems, {PACMI} 2025, Seoul, Republic of Korea, October
                  13-16, 2025},
  pages        = {63--68},
  publisher    = {{ACM}},
  year         = {2025},
  url          = {https://doi.org/10.1145/3766882.3767179},
  doi          = {10.1145/3766882.3767179},
  bibsource    = {dblp computer science bibliography, https://dblp.org}
}

@article{DBLP:journals/corr/abs-2504-05738,
  author       = {Jia Li and
                  Jiacheng Shen and
                  Yuxin Su and
                  Michael R. Lyu},
  title        = {LLM-assisted Mutation for Whitebox {API} Testing},
  journal      = {CoRR},
  volume       = {abs/2504.05738},
  year         = {2025},
  url          = {https://doi.org/10.48550/arXiv.2504.05738},
  doi          = {10.48550/ARXIV.2504.05738},
  eprinttype    = {arXiv},
  eprint       = {2504.05738},
  bibsource    = {dblp computer science bibliography, https://dblp.org}
}

@article{spec-action,
  author       = {Naimeng Ye and
                  Arnav Ahuja and
                  Georgios Liargkovas and
                  Yunan Lu and
                  Kostis Kaffes and
                  Tianyi Peng},
  title        = {Speculative Actions: {A} Lossless Framework for Faster Agentic Systems},
  journal      = {CoRR},
  volume       = {abs/2510.04371},
  year         = {2025},
  url          = {https://doi.org/10.48550/arXiv.2510.04371},
  doi          = {10.48550/ARXIV.2510.04371},
  eprinttype    = {arXiv},
  eprint       = {2510.04371},
  bibsource    = {dblp computer science bibliography, https://dblp.org}
}

@article{spec-edge,
  author       = {Jinwoo Park and
                  Seunggeun Cho and
                  Dongsu Han},
  title        = {SpecEdge: Scalable Edge-Assisted Serving Framework for Interactive
                  LLMs},
  journal      = {CoRR},
  volume       = {abs/2505.17052},
  year         = {2025},
  url          = {https://doi.org/10.48550/arXiv.2505.17052},
  doi          = {10.48550/ARXIV.2505.17052},
  eprinttype    = {arXiv},
  eprint       = {2505.17052},
  bibsource    = {dblp computer science bibliography, https://dblp.org}
}

@inproceedings{DBLP:conf/osdi/BehrensCSBKZ20,
  author       = {Jonathan Behrens and
                  Anton Cao and
                  Cel Skeggs and
                  Adam Belay and
                  M. Frans Kaashoek and
                  Nickolai Zeldovich},
  title        = {Efficiently Mitigating Transient Execution Attacks using the Unmapped
                  Speculation Contract},
  booktitle    = {14th {USENIX} Symposium on Operating Systems Design and Implementation,
                  {OSDI} 2020, Virtual Event, November 4-6, 2020},
  pages        = {1139--1154},
  publisher    = {{USENIX} Association},
  year         = {2020},
  url          = {https://www.usenix.org/conference/osdi20/presentation/behrens},
  bibsource    = {dblp computer science bibliography, https://dblp.org}
}

@inproceedings{DBLP:conf/osdi/ChangG99,
  author       = {Fay W. Chang and
                  Garth A. Gibson},
  editor       = {Margo I. Seltzer and
                  Paul J. Leach},
  title        = {Automatic {I/O} Hint Generation Through Speculative Execution},
  booktitle    = {Proceedings of the Third {USENIX} Symposium on Operating Systems Design
                  and Implementation (OSDI), New Orleans, Louisiana, USA, February 22-25,
                  1999},
  pages        = {1--14},
  publisher    = {{USENIX} Association},
  year         = {1999},
  url          = {https://dl.acm.org/citation.cfm?id=296807},
  bibsource    = {dblp computer science bibliography, https://dblp.org}
}

@article{DBLP:journals/corr/abs-2502-01450,
  author       = {Tianrui Hu and
                  Dimitrios Liakopoulos and
                  Xiwen Wei and
                  Radu Marculescu and
                  Neeraja J. Yadwadkar},
  title        = {Simulating Rumor Spreading in Social Networks using {LLM} Agents},
  journal      = {CoRR},
  volume       = {abs/2502.01450},
  year         = {2025},
  url          = {https://doi.org/10.48550/arXiv.2502.01450},
  doi          = {10.48550/ARXIV.2502.01450},
  eprinttype    = {arXiv},
  eprint       = {2502.01450},
  bibsource    = {dblp computer science bibliography, https://dblp.org}
}

@article{DBLP:journals/corr/abs-2510-26475,
  author       = {Qiaoling Chen and
                  Zijun Liu and
                  Peng Sun and
                  Shenggui Li and
                  Guoteng Wang and
                  Ziming Liu and
                  Yonggang Wen and
                  Siyuan Feng and
                  Tianwei Zhang},
  title        = {ReSpec: Towards Optimizing Speculative Decoding in Reinforcement Learning
                  Systems},
  journal      = {CoRR},
  volume       = {abs/2510.26475},
  year         = {2025},
  url          = {https://doi.org/10.48550/arXiv.2510.26475},
  doi          = {10.48550/ARXIV.2510.26475},
  eprinttype    = {arXiv},
  eprint       = {2510.26475},
  bibsource    = {dblp computer science bibliography, https://dblp.org}
}

@misc{tlt-code,
  title         = {TLT github (FastRL)},
  howpublished = {\burl{https://github.com/mit-han-lab/fastrl}},
  author        = {Qinghao Hu, Shang Yang},
  year          = {2025},
}
}

\clearpage

\end{document}